\documentclass[lettersize,journal]{IEEEtran}
\usepackage{amsmath,amsfonts}
\usepackage{multirow}
\usepackage[ruled,vlined,linesnumbered]{algorithm2e} 
\usepackage{array}
\ifCLASSOPTIONcompsoc
    \usepackage[caption=false, font=normalsize, labelfont=sf, textfont=sf]{subfig}
\else
\usepackage[caption=false, font=footnotesize]{subfig}
\usepackage[colorlinks=true, allcolors=blue]{hyperref}
\usepackage{caption}
\usepackage{textcomp}
\usepackage{stfloats}
\usepackage{url}
\usepackage{verbatim}
\usepackage{graphicx}
\usepackage{cite}
\usepackage{pifont}
\usepackage{booktabs}
\usepackage{amssymb}

\newcommand{\circled}[1]{\ding{\the\numexpr172+#1}}
\AtBeginDocument{%
  \fontdimen2\font=0.21em % 正常间距 (默认大约 0.5em)
  \fontdimen3\font=0.1em  % 可伸展空间
  \fontdimen4\font=0.06em  % 可压缩空间
}

\begin{document}

\title{Morphis: SLO-Aware Resource Scheduling for Microservices with Time-Varying Call Graphs}

\author{
    Yu~Tang,
    Hailiang~Zhao,~\IEEEmembership{Member,~IEEE,}
    Chuansheng~Lu,
    Yifei~Zhang,
    Kingsum~Chow,
    Shuiguang Deng,~\IEEEmembership{Senior~Member,~IEEE},~and
    Rui~Shi% <-this % stops a space
\thanks{
    %%% 类似共一，但是不会在作者上打上标记
    Yu Tang and Hailiang Zhao contribute equally to this work.
}
\thanks{
    Yu Tang, Hailiang Zhao, and Kingsum Chow are with the School of Software Technology, Zhejiang University, Ningbo 315100, China, and also with the Zhejiang Key Laboratory of Digital-Intelligence Service Technology, Hangzhou 310053, China (e-mails: \{y.tang, hliangzhao, kingsum.chow\}@zju.edu.cn).}% <-this % stops a space 
\thanks{
    Chuansheng Lu and Yifei Zhang are with the ByteDance Ltd., Shanghai 800082, China (e-mails: \{ luchuansheng, zhangyifei \}@bytedance.com.).
}
\thanks{Shuiguang Deng is with the College of Computer Science and Technology, Zhejiang University, Hangzhou 310027, China, and also with the Zhejiang Key Laboratory of Digital-Intelligence Service Technology (e-mails: dengsg@zju.edu.cn).}
\thanks{Rui Shi is with the ByteDance Ltd., Shanghai 800082, China (e-mail: shirui@bytedance.com).}
}

        % <-this % stops a space



\maketitle

\begin{abstract}
Modern microservice systems exhibit continuous structural evolution in their runtime call graphs due to workload fluctuations, fault responses, and deployment activities. Despite this complexity, our analysis of over 500,000 production traces from ByteDance reveals a latent regularity: execution paths concentrate around a small set of recurring invocation patterns. However, existing resource management approaches fail to exploit this structure. Industrial autoscalers like Kubernetes HPA ignore inter-service dependencies, while recent academic methods often assume static topologies, rendering them ineffective under dynamic execution contexts.
In this work, we propose Morphis, a dependency-aware provisioning framework that unifies pattern-aware trace analysis with global optimization. It introduces structural fingerprinting that decomposes traces into a stable execution backbone and interpretable deviation subgraphs. Then, resource allocation is formulated as a constrained optimization problem over predicted pattern distributions, jointly minimizing aggregate CPU usage %%% Check: only CPU?
while satisfying end-to-end tail-latency SLOs.
Our extensive evaluations on the \texttt{TrainTicket} benchmark demonstrate that Morphis reduces CPU consumption by 35-38\% compared to state-of-the-art baselines while maintaining 98.8\% SLO compliance.
\end{abstract}


% sections/01-introduction.tex
\section{Introduction}\label{sec:intro}
Cloud-native computing has established microservices as the de facto architecture for modern large-scale distributed systems. By decomposing monolithic applications into fine-grained, independently deployable services, each encapsulating a distinct business capability, organizations gain agility, scalability, and fault isolation~\cite{zhang2025bridge, rzadca2020autopilot}. These services coordinate via network-based remote procedure calls (RPCs) to fulfill end-to-end user requests, collectively forming dynamic call graphs of nontrivial depth and branching complexity. Industry deployments at Meta, Uber, Alibaba, and ByteDance routinely manage tens of thousands of microservices, orchestrating billions of cross-service invocations daily~\cite{zhang2022crisp,huye2023lifting,lu2017imbalance}.

Despite these benefits, microservices introduce a fundamental challenge for resource management: \emph{runtime dependency dynamism}. Unlike static intra-process calls in monoliths, inter-service communication incurs network latency, serialization overhead, and deep, context-sensitive call chains. Critically, the topology of these call graphs is not fixed; it continuously evolves in response to heterogeneous request semantics~\cite{hou2019unleashing, he2022online}, conditional business logic, and fluctuating workloads. Our analysis of production traces from ByteDance reveals that semantically identical user requests often trigger structurally divergent execution paths, varying in depth, fan-out, and service composition, highlighting the inherent unpredictability of real-world microservice ecosystems.
% The fine-grained nature of microservices introduces significant operational complexity. Unlike intra-process function calls in monolithic systems, inter-service communication occurs over the network, resulting in deep call chains and dynamic dependency relationships. 
% A single trace from a production service can involve dozens of interconnected components, with complex branching structures and variable execution paths. The topology of these call graphs changes in real time due to variations in request types~\cite{hou2019unleashing, he2022online}, conditional logic within business workflows, and fluctuating system load. Our analysis of real-world microservice traces reveals that identical user requests may trigger drastically different execution paths across different invocations. These variations manifest in both the length of the call chain and the number of participating services, underscoring the high degree of runtime dynamism inherent in modern microservices.
% This dynamic behavior poses fundamental challenges for performance optimization and resource management. Existing approaches to microservice scheduling often rely on static models of service dependencies, typically represented as fixed directed acyclic graphs~\cite{luo2022erms}. 
This dynamism invalidates two common assumptions in existing schedulers. First, many approaches model dependencies as a static directed acyclic graph (DAG) derived from historical averages~\cite{luo2022erms}, failing to capture contextual adaptations. Second, even dynamic systems often optimize services in isolation, neglecting end-to-end performance coupling: upstream throughput directly shapes downstream load, while downstream bottlenecks induce back-pressure that degrades upstream latency~\cite{li2023topology, ekane2025disc}. Consequently, conventional provisioning strategies suffer from either excessive over-provisioning or SLO violations under realistic workloads.

\begin{figure}[t]
\centering
\subfloat[Total CPU consumption\label{fig:global_opt_cpu_alloc}]{%
    \includegraphics[width=0.45\linewidth]{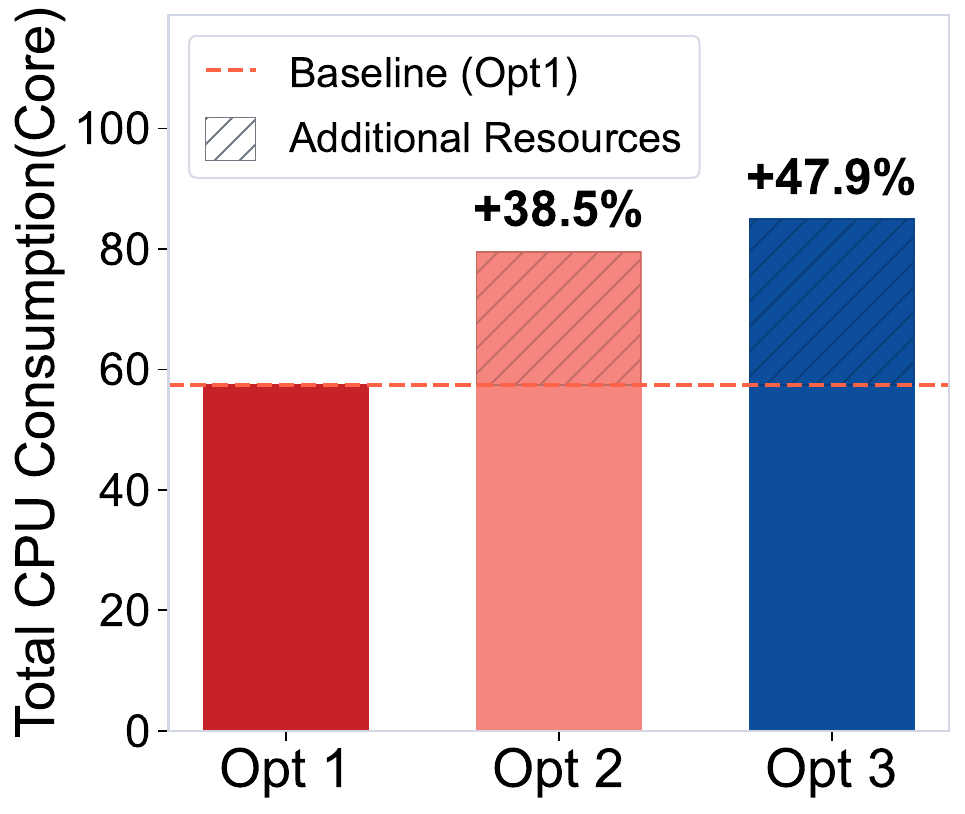}
}
\hfil
\subfloat[CCDF of latency\label{fig:global_opt_latency_ccdf}]{%
    \includegraphics[width=0.45\linewidth]{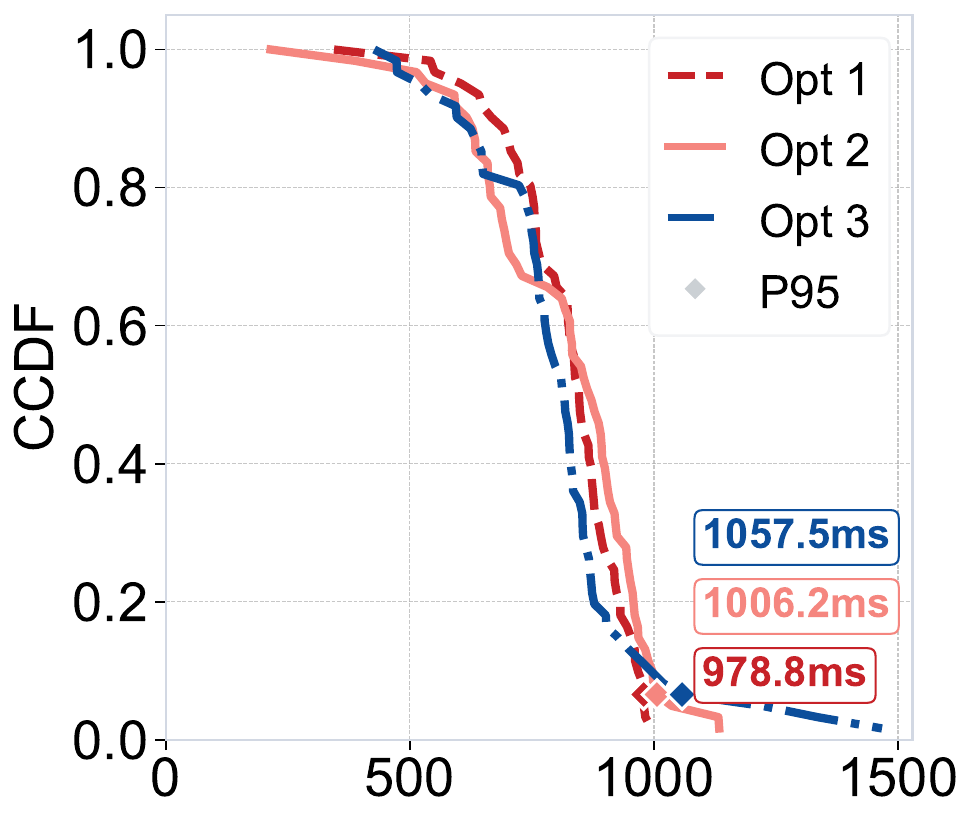}%
}
%%% TODO for ty: 检查CCDF含义是否正确；
\caption{Comparison of three resource provisioning strategies under dynamic workloads. CCDF abbreviates for Complementary Cumulative Distribution Function.} %%% 补充
\label{fig:global_opt_exp_1}
\end{figure}

To quantify the impact of runtime dependency dynamism, we compare three representative strategies under identical workload conditions:  
(i) \textit{Dynamic-Global} (\texttt{Opt1}), which adapts resource allocation based on predicted, evolving call patterns;  
(ii) \textit{Static-Global} (\texttt{Opt2}), which relies on a time-invariant dependency graph derived from historical averages; and  
(iii) \textit{Independent} (\texttt{Opt3}), which performs per-service optimization without dependency awareness.
As shown in Fig.~\ref{fig:global_opt_exp_1}, \textit{Dynamic-Global} achieves substantially higher resource efficiency and tail-latency performance. Specifically, while the baseline deployment consumes approximately 57.4 CPU cores, \textit{Static-Global} requires 38.5\% more resources (79.5 cores), and \textit{Independent} demands 47.9\% additional capacity (84.9 cores). This translates to \textit{Dynamic-Global} delivering 1.4$\times$ and 1.5$\times$ better resource efficiency over \textit{Static-Global} and \textit{Independent}, respectively, while simultaneously improving end-to-end latency compliance (see CCDF in Fig.~\ref{fig:global_opt_latency_ccdf}).
Crucially, despite apparent chaos, our trace analysis reveals a key regularity: the space of dynamic call patterns is both \emph{finite} and \emph{recurring}. Diverse traces are composed of a bounded set of structural motifs, e.g., core workflows, retry cascades, and load-driven parallel expansions, each representing a semantically coherent execution context. This insight enables a paradigm shift: rather than reacting to unbounded graph variations, we can \textit{proactively} manage a compact repertoire of representative patterns.

Leveraging this insight, we present Morphis, a context-aware resource management framework that natively embeds structural understanding into its scheduling intelligence. Morphis operates as a closed-loop system in which structural fingerprinting, designed for distilling dynamic call graphs into a stable execution backbone and interpretable deviation subgraphs, serves as the foundational perception layer. This representation directly fuels a global optimizer that jointly forecasts the prevalence of invocation patterns and allocates resources across services to meet end-to-end SLOs with minimal cost. Our contributions are as follows.
\begin{itemize}
    \item We identify and formalize the problem of \textit{runtime dependency dynamism} in microservice systems, demonstrating through empirical analysis of over 500,000 production traces that static dependency models incur significant resource waste and SLO violations under realistic, evolving workloads.
    \item We introduce \textit{structural fingerprinting} as the core perception mechanism of Morphis: a novel decomposition method that extracts a compact, robust, and semantically meaningful representation of recurring invocation dynamics by separating each trace into a stable performance-critical backbone and context-sensitive deviation subgraphs.
    \item We design Morphis as an integrated framework that unifies structural perception with global optimization. It formulates resource provisioning as a constrained optimization problem over the space of predicted structural fingerprints, enabling coordinated, SLO-aware allocation that adapts to shifting dependency topologies.
    \item We implement Morphis as a Kubernetes-native controller and evaluate it on a large-scale benchmark under diverse workload regimes. Results show it reduces total CPU allocation by 35-38\% compared to state-of-the-art baselines while maintaining 98.8\% end-to-end SLO compliance, validating the efficacy of structure-aware orchestration.
\end{itemize}

The remainder of this paper is organized as follows. Sec. \ref{sec:back} introduces background and motivating examples that highlight the limitations of static dependency models. Sec. \ref{sec:dynasched} introduces the Morphis framework, in which Sec. \ref{subsec:fingerprinting} details our dynamic pattern extraction technique, while Sec. \ref{subsec:opt} describes the resource demand prediction model and global optimization framework. Sec. \ref{sec:eval} presents experiments on real-world microservice clusters under diverse workloads. Sec. \ref{sec:related} reviews related work in microservice management and resource scheduling. Finally, Sec. \ref{sec:conclusion} concludes the paper and discusses future directions.

% The remainder of this paper is organized as follows. Sec. \ref{sec:back} introduces background knowledge and presents motivating examples that illustrate the limitations of static dependency models. Sec. \ref{sec:overview} provides an overview of the Morphis architecture and its workflow. Sec. \ref{sec:pattern} details our dynamic pattern extraction technique. Sec. \ref{sec:opt} describes the resource demand prediction model and the global optimization framework. We present experimental results in Sec. \ref{sec:eval}, 
% evaluating Morphis on real-world microservice clusters under diverse workload scenarios. Sec. \ref{sec:related} reviews related work in microservice management and resource scheduling. Finally, Sec. \ref{sec:conclusion} concludes the paper and discusses future directions.

\section{Background and Motivation}\label{sec:back}

\begin{figure}[t]
    \centering
    % Distribution of span counts per trace in a production cluster, revealing extreme structural variability
    \subfloat[Distribution of span counts. \label{fig:span_nums}]{%
        \includegraphics[width=0.45\linewidth]
        {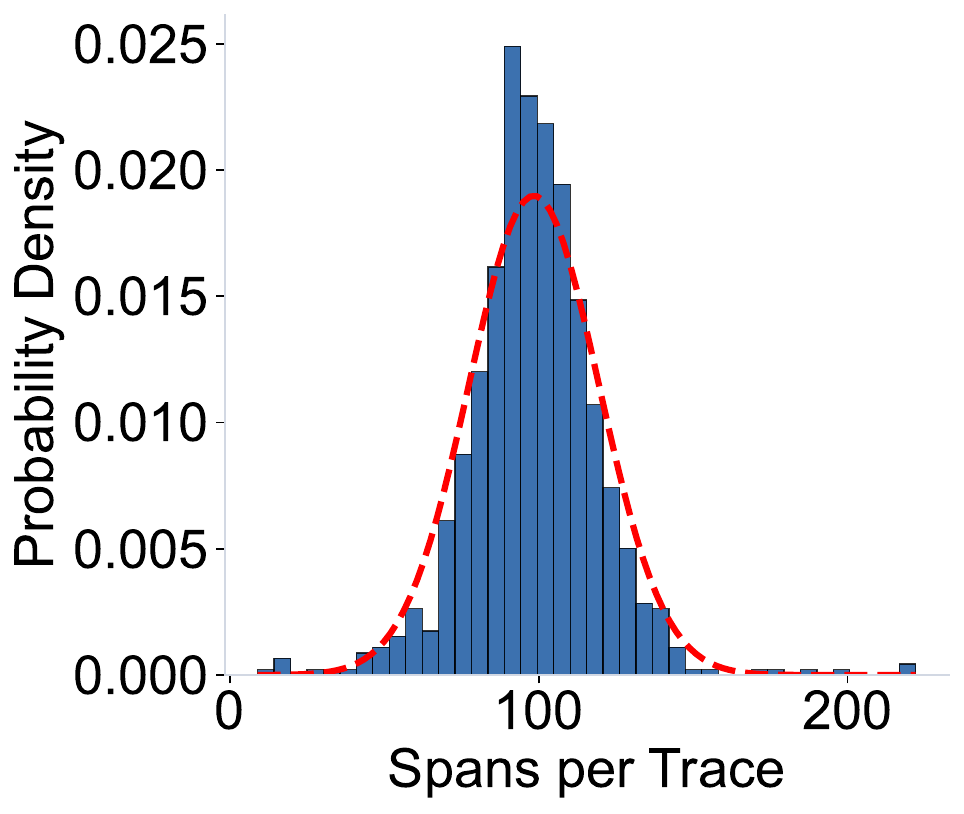}%
    }
    \hfil
    % Complementary CDF of unique call chain coverage: 19 dominant paths account for 90\% of all traces.
    \subfloat[Unique call chain coverage.\label{fig:chain_coverage_ccdf}]{%
        \includegraphics[width=0.45\linewidth]
        {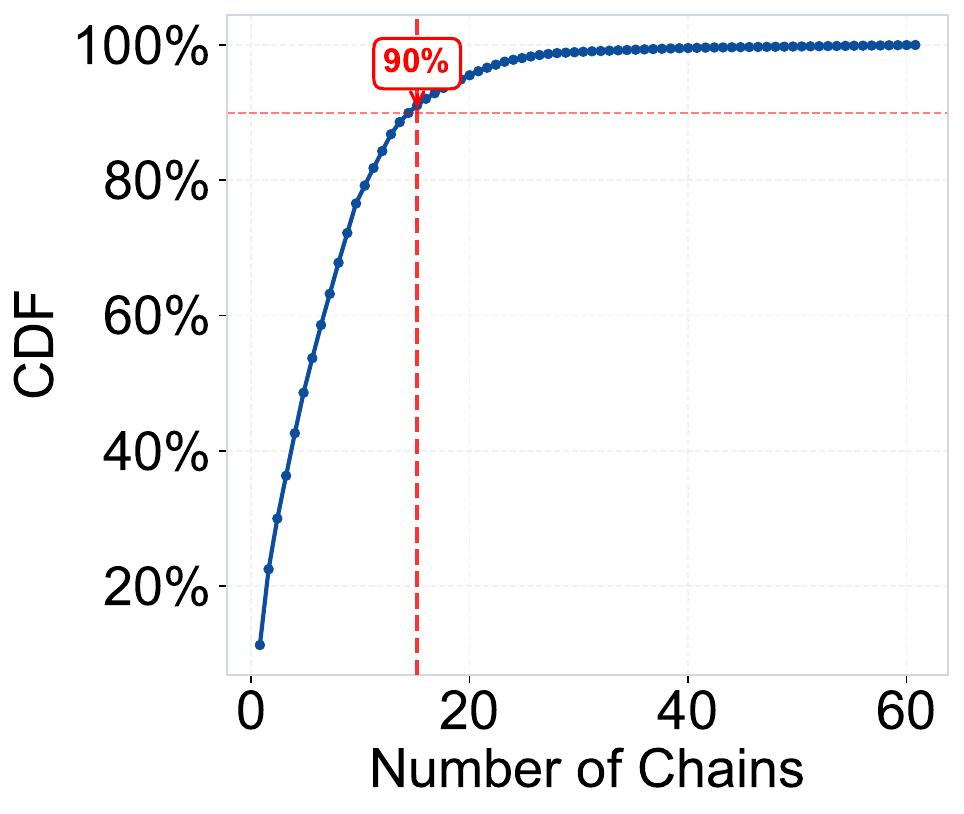}%
    }
    \caption{Dynamic invocation topology in a ByteDance production cluster. (a) Span count distribution per trace, highlighting extreme structural variability. (b) Complementary CDF of call-chain coverage for a key service: 19 dominant patterns account for 90\% of all traces.}
    \label{fig:trace_feature}
\end{figure}

In microservice architectures, a user request typically propagates through a cascade of inter-service invocations, forming an end-to-end execution trace \cite{gan2019open}. The aggregate of such traces defines the system’s runtime dependency structure, commonly abstracted as a directed invocation graph: nodes represent service instances, and edges encode call relationships and data flow.
Critically, unlike the static control flow of monolithic programs, microservice invocation graphs are inherently \textit{dynamic} and \textit{distributed}. Their topology continuously evolves due to service scaling events, deployment rollouts, transient failures, network perturbations, and adaptive resilience mechanisms (e.g., retries, circuit breakers). As illustrated in Fig.~\ref{fig:span_nums}, the number of spans, i.e., individual service calls per trace, in a production cluster varies by over two orders of magnitude between peak and stable periods. This structural volatility renders traditional static dependency models inadequate for performance prediction and resource provisioning.

% \begin{figure}
%     \centering
%     \includegraphics[width=0.6\linewidth]
%     {fig/section2/span_nums_distribution.png}
%     \caption{Trace span count distribution.}
%     \label{fig:span_nums}
% \end{figure}

% \begin{figure}
%     \centering
%     \includegraphics[width=0.6\linewidth]
%     {fig/section2/chain_coverage_ccdf.png}
%     \caption{CCDF of unique call chain coverage.}
%     \label{fig:chain_coverage_cdf}
% \end{figure}

\subsection{Dynamism with Latent Regularity}
Despite the structural instability of microservice traces, our analysis of over 500,000 production traces from ByteDance reveals a latent regularity: trace diversity is not unbounded but concentrates around a small set of recurring \emph{invocation patterns}. As shown in Fig.~\ref{fig:chain_coverage_ccdf}, 19 distinct end-to-end call patterns account for 90\% of all observed traffic, indicating the presence of dominant ``happy paths'' that embody core business logic.
Deviations from these canonical patterns are neither random nor anomalous; they arise from systematic resilience mechanisms triggered by contextual conditions. For example, in a ticket reservation workflow, the \texttt{PreserveTicket} service typically invokes \texttt{Travel}, \texttt{TicketInfo}, and \texttt{Station} sequentially under nominal conditions. However, when \texttt{Station} experiences latency spikes or transient failures, retry logic is activated within the same request context, manifesting in traces as repeated invocations of \texttt{Station} under a single parent span (cf. high-span traces in Fig.~\ref{fig:span_nums}). Similarly, during traffic surges, auto-scaling induces fan-out patterns, e.g., one upstream call dispatches to multiple replicas of \texttt{TicketInfo}. Canary deployments further modulate call destinations as traffic shifts across service versions. Critically, each such deviation follows a structured, semantically meaningful template rather than arbitrary graph mutation.

%%% TODO：文本中需要加一段分析 -- 同一个pattern在不同天展示出来相似的调研频率占比
\begin{figure}[t]
    \centering
    \includegraphics[width=0.9\linewidth]{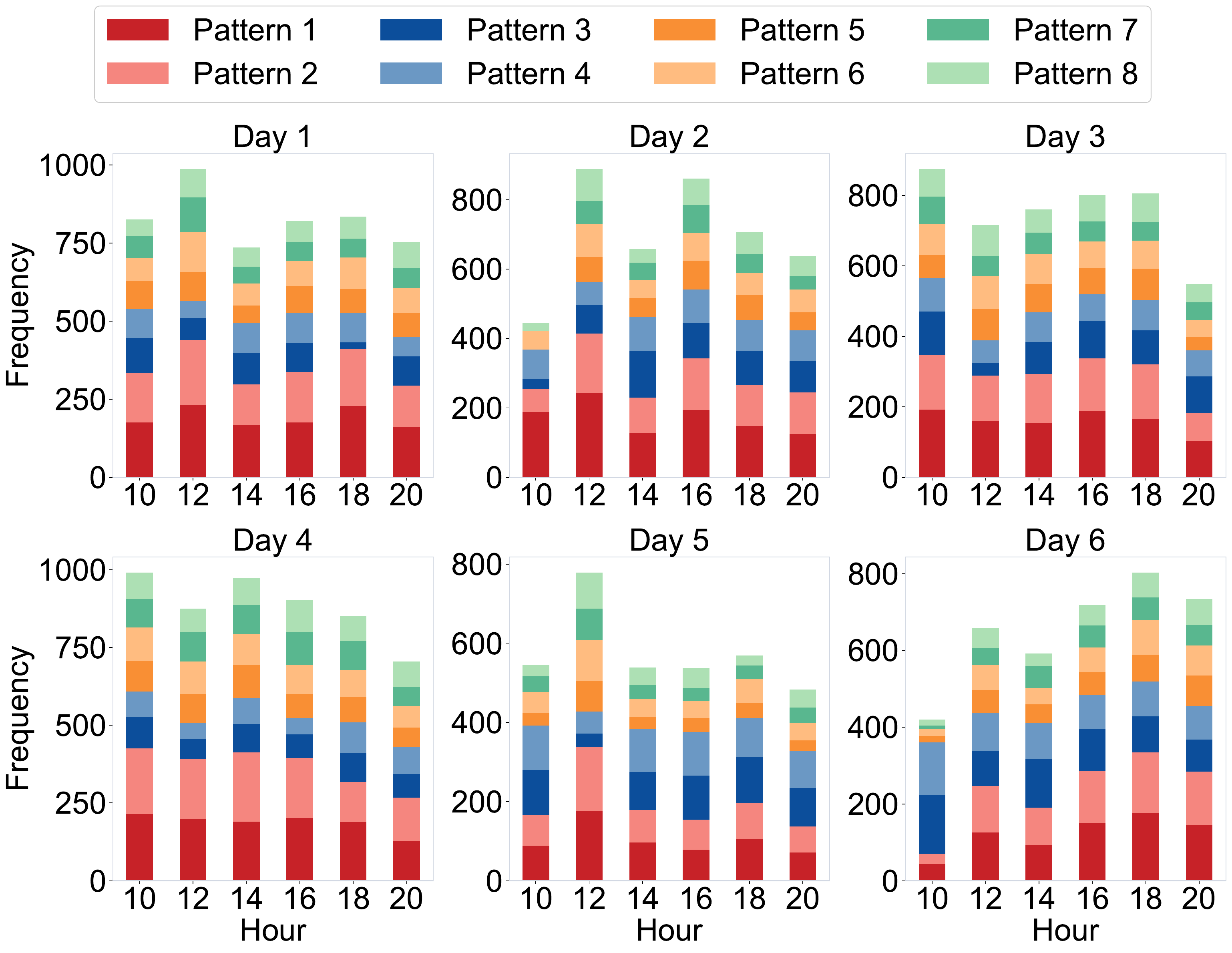}
    \caption{Temporal distribution of dominant call patterns over a 24-hour period, showing strong diurnal periodicity.}
    \label{fig:date_hour_main_pattern}
\end{figure}

Moreover, the prevalence of these invocation patterns exhibits strong temporal coherence. Fig.~\ref{fig:date_hour_main_pattern} shows that dominant patterns recur in alignment with diurnal usage cycles and operational rhythms (e.g., daily peaks, maintenance windows). This demonstrates that call graph evolution is not purely stochastic but governed by latent, learnable dynamics rooted in application semantics and system-level resilience policies. This makes predictive modeling and proactive resource management possible.

\subsection{The Case for Dependency-Aware Resource Management}
The recurring nature of invocation patterns suggests that resource provisioning can and should be both \emph{proactive} and \emph{dependency-aware}. Yet prevailing approaches fall short on both counts. 
In practice, systems like Kubernetes' Horizontal Pod Autoscaler (HPA) scale services in isolation using local metrics, ignoring how upstream scaling propagates load downstream and induces hidden bottlenecks. This reactive, per-service model cannot anticipate structural shifts, e.g., retry cascades or fan-outs, that arise from contextual triggers like failures or traffic surges.
Recent academic efforts have sought to incorporate dependency awareness by leveraging call graphs derived from tracing data. However, many of these methods, e.g., ERMS~\cite{luo2022erms} or Firm~\cite{qiu2020firm}, assume a fixed service topology or optimize individual services under simplified global constraints. In dynamic production environments, where call graphs evolve continuously, such static abstractions quickly become stale, limiting their practical efficacy and adaptability.
As shown in Fig.~\ref{fig:global_opt_exp_1}, this gap has tangible costs: the \textit{Dynamic-Global} strategy that adapts to evolving call patterns reduces resource usage by 38-48\% over both isolated and static-dependency baselines while meeting tail-latency SLOs. This underscores a clear opportunity for a holistic, pattern-driven approach that jointly models structural dynamics and end-to-end performance, which we realize through Morphis.

% Our empirical evaluation validates the necessity of dependency-aware scheduling. As shown in Figure~\ref{fig:global_opt_exp_1}, a strategy that adapts to dynamic call graphs (\texttt{Dynamic-Global}) achieves 38–48\% lower resource consumption than both static-graph and independent baselines while maintaining strict tail-latency SLOs. Even a static dependency model yields measurable gains over isolated optimization, underscoring the inherent value of system-wide coordination.

% Collectively, these findings support our central hypothesis: \emph{the surface-level dynamism of microservice call graphs masks a finite, recurring set of structural motifs that can be systematically extracted and leveraged for proactive, context-aware resource management}. The key challenge lies in (i) identifying these motifs at scale from distributed traces, and (ii) integrating them into low-latency, SLO-constrained optimization loops. In the following sections, we present Morphis, a framework that addresses both challenges through a two-phase pipeline of pattern discovery and global provisioning.
% \section{\texttt{DynaSched}: An Overview}\label{sec:overview}

%%% TODO for ty：拉高左边部分，使其上下高度和右边对齐；不要有$Input$这种文本（不要轻易使用公式环境，普通文本即可）；虚线框实线框太多了，试着去掉一些
\begin{figure*}[t] 
    \centering
    \includegraphics[width=1\textwidth]{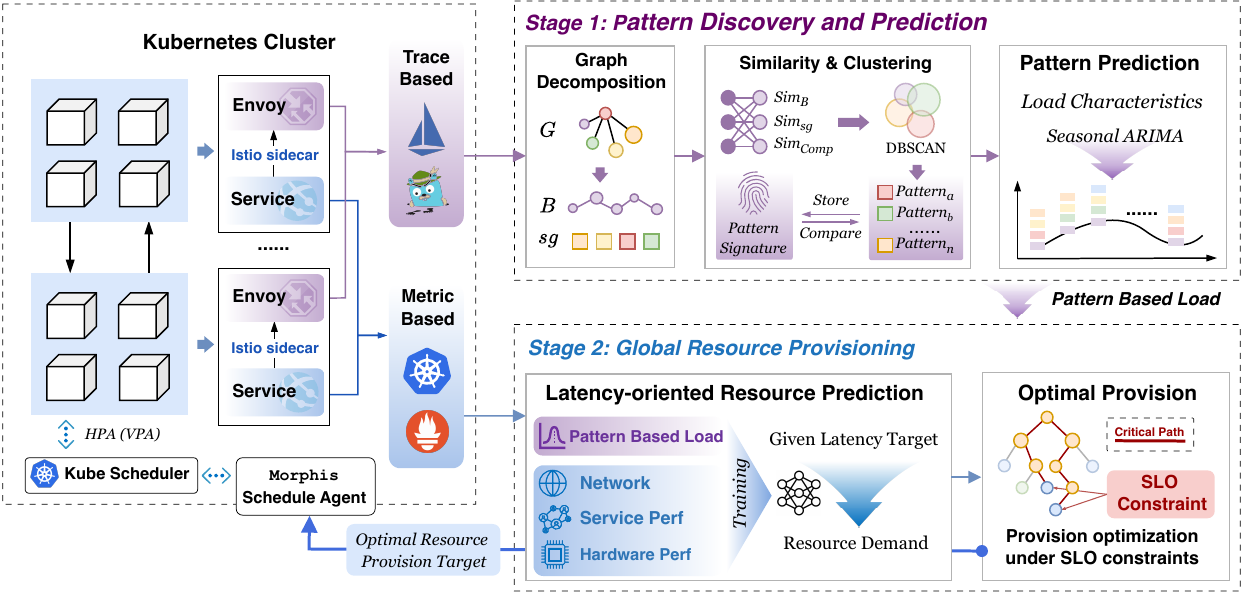} 
    \caption{Workflow of the Morphis framework.}
    \label{fig:system_overview}
\end{figure*}

\section{The Morphis Framework}\label{sec:dynasched}
Morphis is a closed-loop resource management framework composed of two tightly integrated components: (i) structural fingerprinting, which captures the structured evolution of runtime call graphs through interpretable abstractions, and (ii) resource allocation, which leverages these abstractions to synthesize end-to-end SLO-aware provisioning decisions. The pipeline begins with continuous ingestion of distributed traces and system metrics, from which recurring invocation patterns are extracted as structural signatures, each decomposed into a stable execution backbone and context-sensitive deviation subgraphs. These signatures enable short-term forecasting of pattern distributions, which in turn drive a global optimization procedure that computes cost-minimal replica allocations satisfying latency SLOs across all anticipated execution paths. The resulting policy is then safely enacted via orchestration APIs. By grounding resource decisions in the latent regularity of dependency dynamism, Morphis transforms raw observability data into proactive, coordinated control actions.

\subsection{Structural Fingerprinting}
\label{subsec:fingerprinting}

Microservice traces for the same logical request type often exhibit significant structural divergence due to retries, fallbacks, fan-outs, or conditional branches. While these variations reflect intentional resilience mechanisms, treating each trace as unique leads to overfitting, whereas collapsing all traces into a single graph obscures performance-critical deviations. To reconcile stability with discriminability, we introduce \emph{structural fingerprinting}, which represents each trace ensemble as a composition of an invariant backbone and a set of context-dependent deviation subgraphs.

\paragraph{Problem Formulation} 
Let  $ \mathcal{T} = \{ \tau_1, \tau_2, \dots, \tau_n \} $  denote a collection of distributed traces corresponding to requests of the same semantic type (e.g., \texttt{checkout}). Each trace  $ \tau_i $  is modeled as a directed acyclic graph (DAG)  $ \mathcal{G}_i = (\mathcal{V}_i, \mathcal{E}_i, \Phi_i) $ , where $ \mathcal{V}_i $  is the set of service invocation spans, $ \mathcal{E}_i \subseteq \mathcal{V}_i \times \mathcal{V}_i $  encodes parent-child relationships, and $ \Phi_i: \mathcal{V}_i \cup \mathcal{E}_i \to \mathbb{R}^d $  maps nodes and edges to attributes (e.g., latency, error flag, timestamp). Our goal is to construct a structural signature  $ \Psi(\mathcal{T}) := (\mathcal{B}, \mathcal{S}) $  satisfying: (i) \textit{Stability}:  $ \Psi(\mathcal{T}) $  remains consistent under minor, non-semantic variations (e.g., cache hits); (ii) \textit{Discriminability}: Signatures differ when core execution semantics diverge (e.g., normal vs.\ fallback flow); and (iii) \textit{Interpretability}: Components are human-readable and amenable to root-cause analysis. Here,  $ \mathcal{B} $  denotes the execution backbone, and  $ \mathcal{S} = \{ \sigma_1, \dots, \sigma_m \} $  is a set of deviation subgraphs.

\paragraph{Backbone Extraction} 
The backbone $\mathcal{B}$ represents the latency-dominant execution structure that embodies the primary business logic. Unlike naive frequency-based path mining, our approach jointly optimizes \emph{frequency}, \emph{criticality}, and \emph{topological coherence} to identify segments that are not only prevalent but also performance-critical and structurally stable. Given the collection of traces $\mathcal{T}$, we first extract all root-to-leaf invocation paths:  $ \mathcal{P} = \bigcup_{\tau_i \in \mathcal{T}} \texttt{Paths}(\mathcal{G}_i) $ . Each path  $ p \in \mathcal{P} $  is segmented into overlapping  $ k $ -grams (with  $ k=3 $ ), yielding a multiset  $ \Omega $  of local call sequences. This decomposition enables robust pattern matching under minor structural perturbations (e.g., optional logging or cache checks), while preserving sufficient context to distinguish semantically distinct behaviors. For any segment  $ \omega \in \Omega $ , its \emph{support} is defined as the fraction of traces in which  $ \omega $  appears along at least one root-to-leaf path:
\begin{equation}
    s(\omega) = \frac{1}{|\mathcal{T}|} \left| \left\{ \tau_i \in \mathcal{T} \,\middle|\, \exists\, p \in \texttt{Paths}(\mathcal{G}_i) \text{ such that } \omega \sqsubseteq p \right\} \right|,
    \label{eq_supp}
\end{equation}
where $\omega \sqsubseteq p$ denotes that $\omega$ occurs as a contiguous subsequence of $p$. A segment is deemed \emph{frequent} if $s(\omega) \geq \theta$, where $\theta \in (0,1]$ is a tunable threshold chosen to separate core logic from transient deviations (empirically validated in Sec. \ref{sec:eval}). The candidate set is $\Omega_\theta = \{ \omega \in \Omega \mid s(\omega) \geq \theta \}$.

To prioritize performance-relevant segments, we assign each  $ \omega \in \Omega_\theta $  a criticality score  $ \Theta(\omega) \in [0,1] $ :
\begin{equation}
    \Theta(\omega) = \alpha \cdot \frac{1}{|\mathcal{T}|} \sum_{\tau \in \mathcal{T}} \frac{L_\omega(\tau)}{L_{\text{e2e}}(\tau)}
    + (1 - \alpha) \cdot \frac{1}{|\mathcal{T}|} \sum_{\tau \in \mathcal{T}} \frac{|\mathcal{D}_\omega(\tau)|}{|\mathcal{V}(\tau)|},
    \label{eq:criticality}
\end{equation}
where $L_\omega(\tau)$ is the cumulative self-processing latency of nodes in  $ \omega $  within trace $\tau$, $ L_{\text{e2e}}(\tau)$ is the end-to-end latency of $\tau$, and $\mathcal{D}_\omega(\tau)$ is the set of downstream nodes reachable from any node in $\omega$, a proxy for error propagation scope. The parameter  $ \alpha \in [0,1] $  balances latency impact against fault sensitivity.

The backbone $\mathcal{B}$ is constructed by assembling a maximal connected subgraph from $\Omega_\theta$  that maximizes the sum of $\Theta(\cdot)$ scores. This is solved via dynamic programming over the prefix tree of frequent segments: starting from the root, we recursively extend partial backbones by selecting the highest-scoring child segment that maintains topological continuity (i.e., suffix-prefix alignment). To accommodate parallelism (e.g., scatter-gather or fan-out patterns),  $ \mathcal{B} $  is permitted to be a DAG, but its size is constrained to ensure interpretability. Specifically, we limit the number of branching nodes and enforce sparsity by pruning low-$\Theta$ alternatives during assembly. The procedure of backbone extraction is shown in Algorithm \ref{alg:backbone-extraction}.

\begin{algorithm}[htbp]
\caption{Backbone Extraction}
\label{alg:backbone-extraction}
\KwIn{Trace collection  $ \mathcal{T} $ , frequency threshold  $ \theta $ , criticality weight  $ \alpha $ }
\KwOut{Backbone  $ \mathcal{B} = (\mathcal{V}_\mathcal{B}, \mathcal{E}_\mathcal{B}) $ }
 $ \Omega \gets \emptyset $\\
\ForEach{trace  $ \tau \in \mathcal{T} $ }{
    Extract all root-to-leaf paths from  $ \tau $\\
    \ForEach{path  $ p $ }{
        Decompose  $ p $  into overlapping 3-grams and add to  $ \Omega $
    }
}
Compute $s(\omega)$ for each  $ \omega \in \Omega $ by \eqref{eq_supp}\\
$\Omega_\theta \gets \{ \omega \in \Omega \mid s(\omega) \geq \theta \}$\\
\ForEach{$\omega \in \Omega_\theta$}{
    Compute $\Theta(\omega)$ using \eqref{eq:criticality}\\
}
Construct a prefix tree  $ \mathcal{T}_{\text{pref}} $  over  $ \Omega_\theta $\\
Initialize DP table: $F(v) \gets -\infty$ for all nodes  $ v $ in $\mathcal{T}_{\text{pref}}$\\
Set $F(\text{root}) \gets 0$\\
\ForEach{node  $ v $  in BFS order of  $ \mathcal{T}_{\text{pref}} $ }{
    \ForEach{child  $ u $  of  $ v $  corresponding to segment  $ \omega_u $ }{
        \If{ $ F(v) + \Theta(\omega_u) > F(u) $ }{
             $ F(u) \gets F(v) + \Theta(\omega_u) $\\
            Record predecessor of  $ u $  as  $ v $\\
        }
    }
}
Recover optimal path  $ \pi^* $  by backtracking from the node with maximal $F(\cdot)$\\
Convert $\pi^*$ into DAG $\mathcal{B}$ by merging overlapping $k$-grams\\
\KwRet{$\mathcal{B}$}
\end{algorithm}

\paragraph{Deviation Subgraph Extraction}
Given the backbone  $ \mathcal{B} $ , we extract structural deviations as context-dependent subgraphs that capture resilience mechanisms such as retries, circuit breakers, or asynchronous fan-outs. A \emph{divergence point} $v_d$ is any node in $\mathcal{B}$ that has at least one outgoing edge in the original trace graphs $\{\mathcal{G}_i\}$ not present in $\mathcal{B}$. For each $v_d$, we perform a bounded breadth-first search (BFS) over non-backbone edges to delineate the deviation's scope. The traversal terminates upon encountering one of three conditions: (i) a merge point $v_m \in \mathcal{B}$ where execution rejoins the backbone; (ii) a terminal node (e.g., a message queue producer or external stub); or (iii) a maximum depth $\delta_{\max}$ (set to 5 in our implementation) to bound exploration of long-running background tasks.

To ensure disjoint coverage and eliminate ambiguity in nested or overlapping deviations, we assign each non-backbone node to exactly one divergence point via an \emph{ownership map}. Specifically, for any node $u \notin \mathcal{V}(\mathcal{B})$, let $\Pi(u)$ denote the set of divergence points from which $u$  is reachable through non-backbone paths. We assign $u$ to the divergence point $ v_d^* \in \Pi(u)$ that appears earliest in the topological order of $\mathcal{B}$. This guarantees a unique, interpretable attribution of every deviation fragment.
Each extracted deviation subgraph  $ \sigma $  is represented as a structured tuple:
\begin{equation}
    \sigma = \big( \mathcal{V}_\sigma, \mathcal{E}_\sigma, v_d, v_m, \boldsymbol{\phi}_\sigma \big),
    \label{eq:subgraph_tuple}
\end{equation}
where  $ \mathcal{V}_\sigma $  and  $ \mathcal{E}_\sigma $  are the nodes and edges of the induced subgraph,  $ v_d $  is the divergence point,  $ v_m \in \mathcal{V}(\mathcal{B}) \cup \{\varnothing\} $  is the merge point (or null if none exists), and  $ \boldsymbol{\phi}_\sigma \in \mathbb{R}^5 $  is a feature vector encoding five key attributes:
\begin{equation}
    \boldsymbol{\phi}_\sigma = \Big[ \text{depth},\ \text{width},\ |\mathcal{V}_\sigma|,\ \bar{p}_{\text{exec}},\ \bar{p}_{\text{fail}} \Big]^\textrm{T},
\end{equation}
with $\bar{p}_{\text{exec}}$ and $\bar{p}_{\text{fail}}$ denoting the average execution probability and failure rate of the subgraph across traces in $\mathcal{T}$, respectively. The complete structural signature is then defined as  $\Psi(\mathcal{T}) := (\mathcal{B}, \mathcal{S})$, where $\mathcal{S} = \{ \sigma_1, \dots, \sigma_m \}$ is the set of all extracted deviation subgraphs.

\paragraph{Similarity Computation}
%%% TODO：这一段内容过于偏工程，需要看看如何能加一些理论分析
To enable pattern clustering and forecasting, we define a similarity metric between two structural signatures  $ S_1 = (\mathcal{B}_1, \mathcal{S}_1) $  and  $ S_2 = (\mathcal{B}_2, \mathcal{S}_2) $ . This metric operates at three hierarchical levels, i.e., backbone alignment, subgraph correspondence, and service composition, and fuses them into a unified score adapted to the target system's operational characteristics.
Backbone similarity accounts for both structural overlap and semantic equivalence. Let  $ \texttt{LCS}(\mathcal{B}_1, \mathcal{B}_2) $  denote the longest common subsequence of service invocations between the two backbones, where services are considered matching if they share identical names or belong to the same functional category (e.g., ``auth-service'' and ``oauth-provider''). Define the match quality of aligned service pair  $ (s_1^i, s_2^i) $  as:
%%% TODO: Why 1.0 and 0.8? Need a theory?
\begin{equation}
    M(s_1^i, s_2^i) =
    \begin{cases}
        1.0 & \text{if } s_1^i = s_2^i, \\
        0.8 & \text{if functionally equivalent}, \\
        0   & \text{otherwise}.
    \end{cases}
    \label{eq:match_func}
\end{equation}
The backbone similarity is then:
\begin{equation}
    \text{SIM}_{\text{BG}}(\mathcal{B}_1, \mathcal{B}_2) = 
    \frac{2 \cdot \big|\texttt{LCS}(\mathcal{B}_1, \mathcal{B}_2)\big|}{|\mathcal{B}_1| + |\mathcal{B}_2 |}
    \cdot \prod_{i=1}^{|\texttt{LCS}|} M \big(s_1^i, s_2^i \big).
    \label{eq:sim_bg}
\end{equation}

Subgraph similarity requires establishing cross-signature correspondence. We formulate this as a constrained bipartite matching problem between $\mathcal{S}_1$ and $\mathcal{S}_2$. A pair $(\sigma_1, \sigma_2)$ is eligible for matching only if: (i) their divergence points occur at comparable relative positions along their backbones (measured by normalized index distance $< 0.2$); and (ii) their pattern types (classified via $\boldsymbol{\phi}_\sigma$ using a lightweight decision tree) are consistent (e.g., both represent retries). For eligible pairs, we compute hybrid similarity:
\begin{align}
    \text{SIM}_{\text{SG}}(\sigma_1, \sigma_2) = 
    \beta \left(1 - \frac{\text{GED}(\sigma_1, \sigma_2)}{N_{\max}}\right) \nonumber\\
    + (1 - \beta) \cos(\boldsymbol{\phi}_{\sigma_1}, \boldsymbol{\phi}_{\sigma_2}),
    \label{eq:sim_sg}
\end{align}
where  $ \text{GED}(\cdot,\cdot) $  is an approximate graph edit distance,  $ N_{\max} = \max(|\mathcal{V}_1| + |\mathcal{E}_1|, |\mathcal{V}_2| + |\mathcal{E}_2|) $ , and  $ \beta = 0.5 $  balances structure and features. The optimal assignment  $ \mathcal{M}^* $  is obtained via the Hungarian algorithm, and the aggregate subgraph similarity is:
\begin{equation}
    \text{SIM}_{\text{subgraphs}} = \frac{1}{\max(|\mathcal{S}_1|, |\mathcal{S}_2|)} \sum_{(\sigma_1, \sigma_2) \in \mathcal{M}^*} \text{SIM}_{\text{SG}}(\sigma_1, \sigma_2).
    \label{eq:sim_subgraphs}
\end{equation}
Finally, service composition similarity is measured via Jaccard index on service name sets, smoothed by functional categories to tolerate naming variance. The overall similarity is a convex combination:
\begin{equation}
    \text{SIM}(S_1, S_2) = w_b \cdot \text{SIM}_{\text{BG}} + w_s \cdot \text{SIM}_{\text{subgraphs}} + w_c \cdot \text{SIM}_{\text{comp}},
    \label{eq:overall_sim}
\end{equation}
where weights  $ (w_b, w_s, w_c) $  are dynamically inferred during an initialization phase by analyzing the variance of backbone vs. subgraph structures across historical traces. In synchronous, latency-sensitive workflows,  $ w_b $  dominates; in event-driven systems,  $ w_s $  increases accordingly. This adaptive fusion ensures the fingerprint remains discriminative across diverse microservice architectures.

\subsection{Resource Modeling and Optimization}\label{subsec:opt}
\begin{figure}[t]
    \centering
    \includegraphics[width=0.8\linewidth]{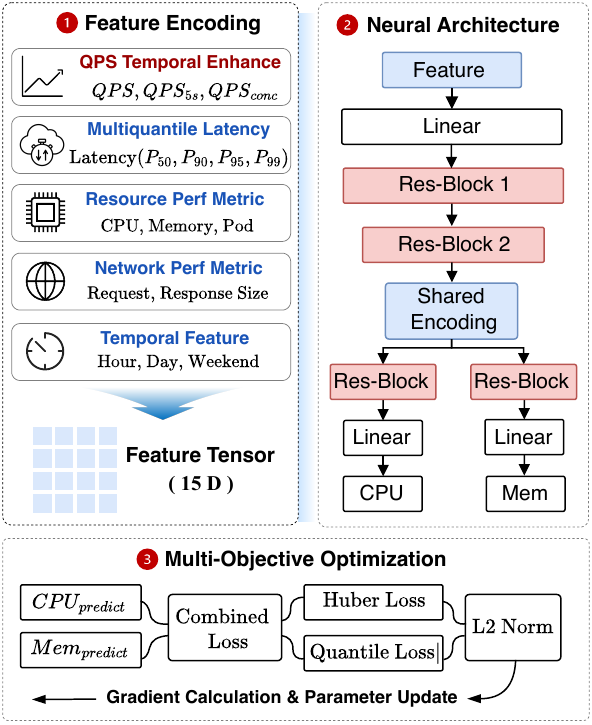}
    \caption{Overview of the resource prediction and optimization pipeline.}
    \label{fig:resource_prediction}
\end{figure}

Proactive resource provisioning in microservice systems demands a holistic framework that bridges fine-grained performance modeling with system-wide optimization. While per-service resource predictors offer local insights, their isolated application fails to honor end-to-end latency objectives shaped by complex invocation topologies. To address this, we present a two-stage optimization pipeline: (i) a latency-aware neural model that predicts CPU and memory consumption under self-processing latency targets, and (ii) a global combinatorial optimizer that allocates these targets across services to minimize total cost while satisfying path-level SLOs. The overall workflow is illustrated in Fig. \ref{fig:resource_prediction}.

%%% TODO for ty：第二部分神经网络架构中，箭头替换成细的黑色箭头（第三部分中的箭头）；字体统一为无衬线字体；公式/符号则使用 latex 字体

\begin{figure}
    \centering
    \includegraphics[width=0.8\linewidth]{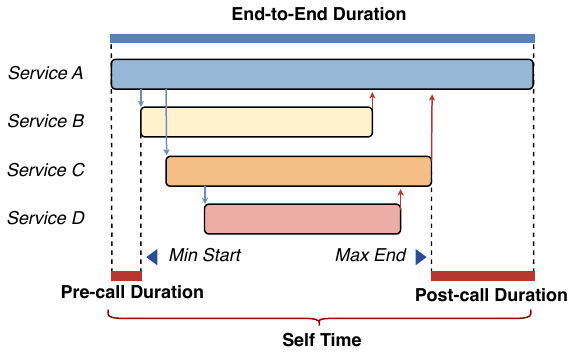}
    \caption{Decomposition of end-to-end latency into self-processing and downstream components. Only self-processing time is used for local resource prediction.}
    \label{fig:self_time}
\end{figure}

\paragraph{Latency-Aware Resource Modeling}
Conventional latency metrics, such as total response time reported by service meshes like Istio \cite{Istio}, include downstream call durations, conflating local computation with external dependencies. This renders them unsuitable for capacity planning, as resource pressure correlates primarily with \emph{self-processing latency}, defined as the time a service spends executing its own logic, excluding child invocations. As shown in Fig. \ref{fig:self_time}, decomposing end-to-end traces into self-time components enables accurate modeling of local resource demands.

For each service $s_i$, we train a deep neural network that maps a contextual feature vector to predicted CPU and memory usage under a given self-latency target $\tau_i$. The input features capture workload dynamics (instantaneous QPS, 5-second sliding-window QPS, concurrent requests), tail latency behavior (P50-P99 self-times), and system pressure (CPU/memory utilization, network I/O). The architecture employs stacked residual blocks \cite{zhang2017residual} for stable deep representation learning, augmented with an attention mechanism that dynamically reweights features based on operational context, e.g., emphasizing latency during degradation and throughput during steady state.
The model adopts a multi-task design with a shared backbone and two task-specific heads for CPU and memory prediction, enabling knowledge transfer while preserving modality-specific dynamics. Training minimizes a composite loss combining Huber loss (for robustness to outliers) and quantile loss (to capture conditional distributions). Inference yields a cost function:
\begin{equation}
    C(s_i,q_i^t,\tau_i)=\alpha\cdot\text{CPU}(s_i,q_i^t,\tau_i)+\beta\cdot\text{Memory}(s_i,q_i^t,\tau_i),
    \label{eq:cost_function}
\end{equation}
where $q_i^t$ is the predicted QPS for service  $ s_i $  in time window $t$, and  $ \alpha,\beta $  are configurable cost weights reflecting infrastructure pricing.

%%% TODO for ty: 字体全部保持统一，使用无衬线字体

\paragraph{Pattern-Based Load Forecasting}
To compute $q_i^t$, we avoid per-service forecasting, which vulnerable to error propagation under shifting call compositions. Instead, we forecast at the level of \emph{structural fingerprints}  $ \Psi_j=(\mathcal{B}_j,\mathcal{S}_j) $  introduced in Sec. \ref{subsec:fingerprinting}. Each fingerprint corresponds to a distinct logical request type (e.g., ``checkout with fallback''), and its frequency determines the realized load on constituent services.

Let $\mathcal{K}=\{\Psi_1,\dots,\Psi_m\}$ be the set of observed patterns for a given entrypoint. For each $\Psi_j$, we forecast its incoming rate $q_j^t$ using a Seasonal ARIMA with eXogenous regressors (SARIMAX):
\begin{equation}
    q_j^t=\beta_j^\top x_t+u_t^{(j)},
    \label{eq:sarimax_load}
\end{equation}
where $x_t=[(x_t^{(h)})^\top,(x_t^{(d)})^\top,x_t^{(s)}]^\top$ encodes hour-of-day, day-of-week, and special-day indicators, and  $ u_t^{(j)} $  follows a SARIMA process capturing autocorrelation and weekly seasonality. Predictions are refreshed every 5 minutes via a rolling short-horizon strategy to limit error accumulation.
Per-service QPS is then derived by propagating pattern rates through the call graph:
\begin{equation}
    q_i^t=\sum_{j=1}^m q_j^t\cdot\eta_{j,i},
    \label{eq:qps_propagation}
\end{equation}
where $\eta_{j,i}$ is the expected number of invocations of service $s_i$ per request of pattern $\Psi_j$, precomputed from historical traces. This dependency-aware propagation ensures workload estimates reflect actual execution semantics.

\paragraph{Global Resource Allocation Optimization}
Given predicted workloads $\{q_i^t\}$ and the cost model $C(\cdot)$, we solve a constrained optimization problem to assign self-latency targets $\tau_i$ that minimize total resource cost while meeting end-to-end SLOs. Let $S=\{s_1,\dots,s_n\}$ denote the service set, and let $P=\{p_1,\dots,p_m\}$ be the collection of critical execution paths, each $p_j$ being a sequence of services derived from backbones and high-impact deviation subgraphs. The optimization problem is:
\begin{equation}
\begin{aligned}
\min_{\boldsymbol{\tau}} &\quad \sum_{i=1}^{n} C(s_i,q_i^t,\tau_i) \\
\text{s.t.} &\quad \sum_{s_i\in p_j}\tau_i\leq T_{\text{e2e}},\quad \forall p_j\in P, \\
            &\quad \tau_i\in\Omega_i,\quad \forall s_i\in S,
\end{aligned}
\label{eq:global_opt}
\end{equation}
where  $ T_{\text{e2e}} $  is the global latency budget, and  $ \Omega_i $  is the feasible SLO space for service  $ s_i $ .

The cost function  $ C(\cdot) $  is a black-box neural model, rendering continuous optimization intractable. We therefore discretize  $ \Omega_i $  into a finite set:
\begin{equation}
    \Omega_i=\{\tau_i^{\min},\tau_i^{\min}+\Delta\tau,\dots,\tau_i^{\max}\},
\end{equation}
where bounds are derived from empirical latency percentiles (e.g., P1-P99), ensuring candidate SLOs remain operationally realistic.
To avoid repeated neural inference during optimization, we precompute a lookup cache:
\begin{equation}
    \mathcal{R}=\left\{(s_i,q,\tau)\mapsto\big(\text{cpu}_{i,q,\tau},\text{mem}_{i,q,\tau}\big)\;\middle|\;
    \begin{array}{l}
        s_i\in S,\\
        q\in Q_i,\\
        \tau\in\Omega_i
    \end{array}
    \right\},
    \label{eq:cache}
\end{equation}
where $Q_i$ is the set of observed load levels for $s_i$. Precomputation is parallelized across workers, each holding a local model copy, reducing per-query latency from milliseconds to microseconds.

We solve \eqref{eq:global_opt} using a genetic algorithm (GA) tailored to the discrete, constrained search space. Each chromosome  $ \mathbf{x}=[\tau_1,\dots,\tau_n] $  encodes a complete SLO assignment. The fitness function balances cost and constraint violation:
\begin{equation}
    f(\mathbf{x})=\sum_{i=1}^{n} C(s_i,q_i^t,\tau_i)+\lambda\cdot V(\mathbf{x}),
    \label{eq:fitness}
\end{equation}
where $V(\mathbf{x})=\sum_{p_j\in P}\max\big(0,\sum_{s_i\in p_j}\tau_i-T_{\text{e2e}}\big)$ is the total path-wise latency violation, and $\lambda$ is a penalty coefficient.
The GA uses tournament selection (size 3), two-point crossover (probability 0.7), and service-aware mutation (per-gene probability 0.1, resampling from $\Omega_i$). Elitism preserves the best solution across generations. The algorithm runs for up to 70 generations, with early stopping if fitness stagnates for 10 consecutive generations. In practice, it converges within seconds, yielding near-optimal allocations that respect both cost and SLO constraints.

\begin{figure}
    \centering
    \includegraphics[width=1\linewidth]{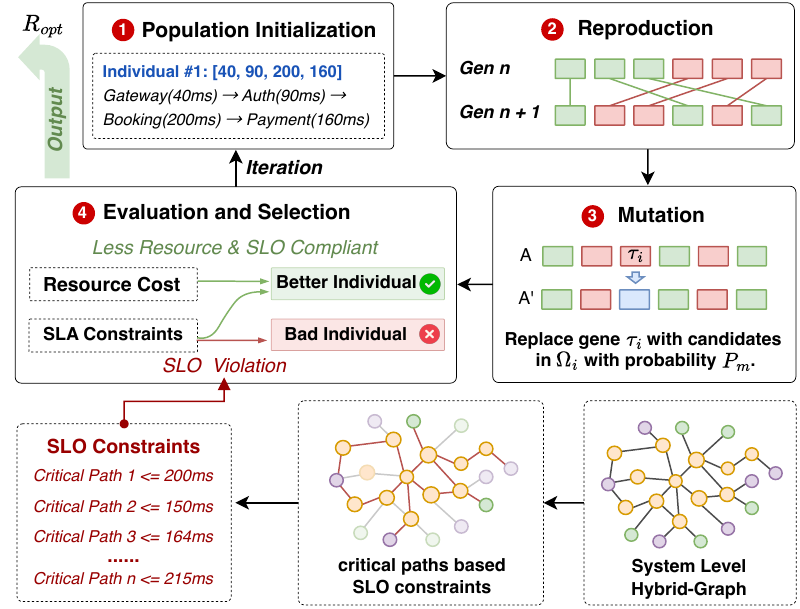}
    \caption{Workflow of the global resource allocation optimization process, including preprocessing, per-window optimization, and genetic search.}
    \label{fig:global_opt_func}
\end{figure}

The optimization workflow is depicted in Fig. \ref{fig:global_opt_func}. The complete procedure is summarized in Algorithm~\ref{alg:global-allocation}. Note that all steps are grounded in the structural fingerprinting framework, ensuring semantic consistency between prediction and control.

\begin{algorithm}[htbp]
\caption{Global Resource Optimization}
\label{alg:global-allocation}
\KwIn{Time horizon  $ \mathcal{T} $ , service set  $ S $ , historical traces  $ H $ , global SLO  $ T_{\text{e2e}} $ }
\KwOut{Optimal allocation plan  $ \{\boldsymbol{\tau}^t\}_{t \in \mathcal{T}} $ }
\ForEach{ $ s_i \in S $ }{
    Estimate empirical self-latency bounds  $ (\tau_i^{\min}, \tau_i^{\max}) $  from  $ H $\\
    Discretize into  $ \Omega_i \gets \{\tau_i^{\min}, \tau_i^{\min} + \Delta\tau, \dots, \tau_i^{\max}\} $\\
}
 $ \mathcal{R} \gets \emptyset $\\
\ForEach{ $ s_i \in S $ , observed load  $ q \in Q_i $ ,  $ \tau \in \Omega_i $ }{
     $ (c,m) \gets \text{NeuralModel}(s_i, q, \tau) $\\
     $ \mathcal{R} \gets \mathcal{R} \cup \{(s_i,q,\tau) \mapsto (c,m)\} $\\
}
\ForEach{time window  $ t \in \mathcal{T} $ }{
    % Forecast pattern rates and propagate to per-service QPS
    \ForEach{structural fingerprint  $ \Psi_j $ }{
        Forecast incoming rate  $ q_j^t $  using SARIMAX\\
    }
    \ForEach{ $ s_i \in S $ }{
        Compute  $ q_i^t \gets \sum_j q_j^t \cdot \eta_{j,i} $\\
    }
    % Enumerate critical paths
     $ P^t \gets \bigcup_j \big(\text{Backbone}(\Psi_j) \cup \text{HighImpactSubgraphs}(\Psi_j)\big) $\\
    % Solve GA for optimal  $ \tau $ 
     $ \boldsymbol{\tau}^t \gets \text{GeneticSearch}\big(S, P^t, \{q_i^t\}, \mathcal{R}, T_{\text{e2e}}\big) $\\
    Store  $ \boldsymbol{\tau}^t $  in allocation plan\\
}
\KwRet{ $ \{\boldsymbol{\tau}^t\}_{t \in \mathcal{T}} $ }
\end{algorithm}

\section{Evaluation}\label{sec:eval}
We conduct a comprehensive evaluation of our resource optimization framework across three key dimensions: (i) the accuracy and robustness of call pattern recognition, (ii) the precision of the resource demand prediction module compared to state-of-the-art models, and (iii) the end-to-end performance of the full system against widely adopted auto-scaling baselines under realistic microservice workloads. Our evaluation encompasses both component-level metrics and system-level outcomes in terms of quality of service (QoS) and resource cost.

\subsection{System Implementation}
We implement Morphis as a Kubernetes-native controller written in Go. The system integrates seamlessly with standard observability and orchestration infrastructure: distributed traces are collected via Jaeger~\cite{Jaeger}, container-level CPU and memory metrics are pulled from Prometheus~\cite{Prometheus}, and resource policies, e.g., replica counts and resource requests, are applied through the Kubernetes API server using native Deployment and Horizontal Pod Autoscaler (HPA) abstractions.\footnote{In Kubernetes, a Deployment manages a set of Pods to run a stateless application workload. A HPA automatically updates a workload resource (such as a Deployment or StatefulSet), with the aim of automatically scaling capacity to match demand.} All components run as pods within a dedicated control-plane namespace, ensuring isolation, security, and horizontal scalability.

The optimizer employs a genetic algorithm enhanced with warm-starting from solutions computed in the previous time window, thereby meeting sub-minute decision latency requirements even in production-scale clusters. Internally, the system maintains a precomputed resource cache  $ \mathcal{R} $  (as defined in \eqref{eq:cache}) to accelerate neural inference during optimization. The neural predictor is implemented using PyTorch and served via TorchServe, while the SARIMAX forecaster leverages the statsmodels library.\footnote{\url{https://www.statsmodels.org/stable/index.html}} All modules communicate through gRPC, and shared state is managed via an in-memory Redis instance for low-latency access.

%%% TODO for ty: 检查超参数设定是否全面、是否正确
% \textcolor{red}{
Key hyperparameters are configured as follows: the structural fingerprinting window spans 5 minutes, aligning with typical SLO monitoring intervals.\footnote{In our testbed, we find that structural patterns evolve slowly relative to the 5-minute window. Rapid topology changes (e.g., full canary rollout) may require shorter windows or online clustering updates.} 
% The pattern clustering uses DBSCAN with $\epsilon = 0.3$ and $\texttt{min\_samples} = 4$ to balance granularity and robustness;
The pattern clustering uses DBSCAN with a dynamic search for the optimal $\epsilon$ value and setting $\texttt{min\_samples} = 4$ to minimize classification noise and balance clustering granularity with robustness;
the neural predictor is a 3-layer MLP with 128 hidden units per layer, trained with Adam ($\texttt{lr}=10^{-3}$, weight decay  $ =10^{-4}$); the SARIMAX model uses order $(1,1,1)$ and seasonal order $ (1,0,1,7)$ to capture daily periodicity; the genetic algorithm runs for 30 generations with a population size of 50, crossover rate 0.8, and mutation rate 0.1; and the SLO violation penalty coefficient $\lambda$ in the objective function is set to 100 based on grid search over validation workloads.
% }

\subsection{Experimental Setup}
The experimental setup for our evaluation is described as follows.

\paragraph{Benchmark}
We use \texttt{TrainTicket} \cite{zhou2018benchmarking} to benchmark the experiments. \texttt{TrainTicket} is an open-source railway ticket booking system composed of 41 heterogeneous microservices, including both stateful components (e.g., user profile management, order processing) and stateless services (e.g., search, authentication). It has rich and dynamic call graphs, which are characterized by cascading invocations, fan-out patterns, and conditional branching. This makes it particularly well-suited for evaluating context-aware resource optimization strategies.

\paragraph{Cluster setup and self-collected trace}
Our testbed consists of a Kubernetes cluster deployed across three physical nodes, each equipped with a 64-core CPU and 128\,GB of RAM, interconnected via a 10\,Gbps network. We execute a continuous 40-hour stress test covering three primary transaction flows: \texttt{QueryTicket}, \texttt{BookTicket}, and \texttt{Pay}. During this period, we collect fine-grained telemetry data, including per-service latency, throughput (QPS), resource utilization, and trace-level execution paths, resulting in a total of \textbf{973,602} valid data samples. The dataset is randomly shuffled and split into a 70\% training set and a 30\% validation set for model training and hyperparameter tuning. This partitioning strategy preserves both the temporal dynamics and the distributional characteristics of real-world traffic.

\begin{figure}
    \centering
    \includegraphics[width=1\linewidth]{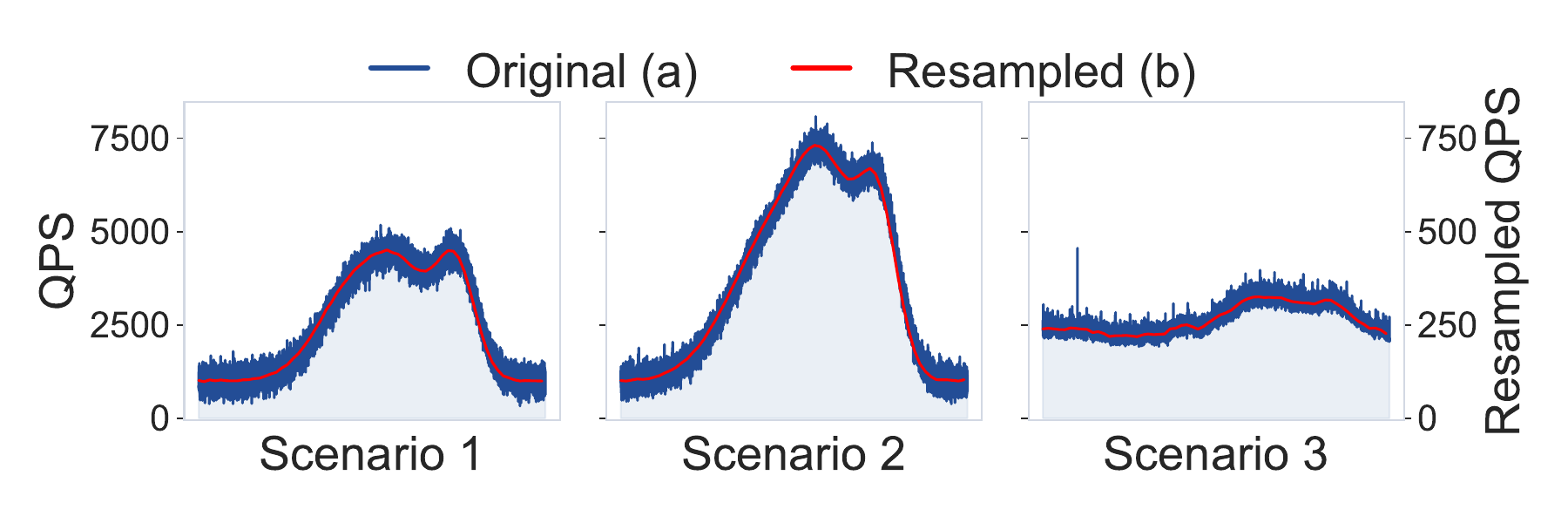}
    \caption{Production-derived workload patterns used in evaluation: (a) original traffic trace, (b) resampled and normalized load shapes for Gradual Rise, High Sustained Load, and Spike-and-Plateau scenarios.}
    \label{fig:exp_load_shape}
\end{figure}

\paragraph{Workloads} To evaluate system adaptability, we employ three canonical workload patterns resampled from production traces: (i) \textit{Gradual Rise}, featuring a smooth QPS increase to assess proactive scaling capabilities; (ii) \textit{High Sustained Load}, representing prolonged peak intensity to evaluate stability and cost efficiency; and (iii) \textit{Spike and Plateau}, involving a sudden surge followed by sustained load to challenge system responsiveness and consolidation behavior.
Each pattern is normalized and compressed into a 60-minute experimental window while preserving its relative intensity and temporal structure. The original and resampled load profiles are visualized in Fig. \ref{fig:exp_load_shape}. These representative scenarios enable rigorous assessment across diverse scaling dynamics and resource allocation trade-offs.

%%% TODO for ty：这里的 (a) (b) 在哪里

\paragraph{Baselines} We compare our approach against three representative auto-scaling strategies drawn from both academia and industry:

\begin{itemize}
    \item Derm~\cite{chen2024derm}: A learning-based method that models dynamic call dependencies to predict resource needs. Like our approach, Derm incorporates service interaction patterns but does not perform global SLO-aware optimization.
    \item PBScaler~\cite{xie2024pbscaler}: A performance bottleneck-aware scaler that detects latency degradation and triggers scaling on bottleneck services. It employs queueing theory to identify bottlenecks but lacks proactive prediction and inter-service coordination.
    \item HPA~\cite{kubernetes_hpa}: The default Kubernetes Horizontal Pod Autoscaler, which scales pods based on observed CPU or memory utilization (threshold: 80\%). It represents a widely deployed industry standard but is purely reactive and agnostic to end-to-end SLOs.
\end{itemize}

All baselines are configured with their recommended parameters and integrated into the same Kubernetes environment to ensure a fair and reproducible comparison.

\subsection{Invoke Pattern Recognition}
We evaluate the effectiveness and efficiency of our invocation pattern recognition module across multiple production services and the \texttt{TrainTicket} benchmark, spanning diverse business domains including AI inference, comment processing, and e-commerce transaction systems. These domains exhibit significantly different call graph characteristics in terms of scale, depth, and branching complexity, thereby enabling a robust assessment of generalization capability.

\begin{figure}
    \centering
    \includegraphics[width=0.7\linewidth]{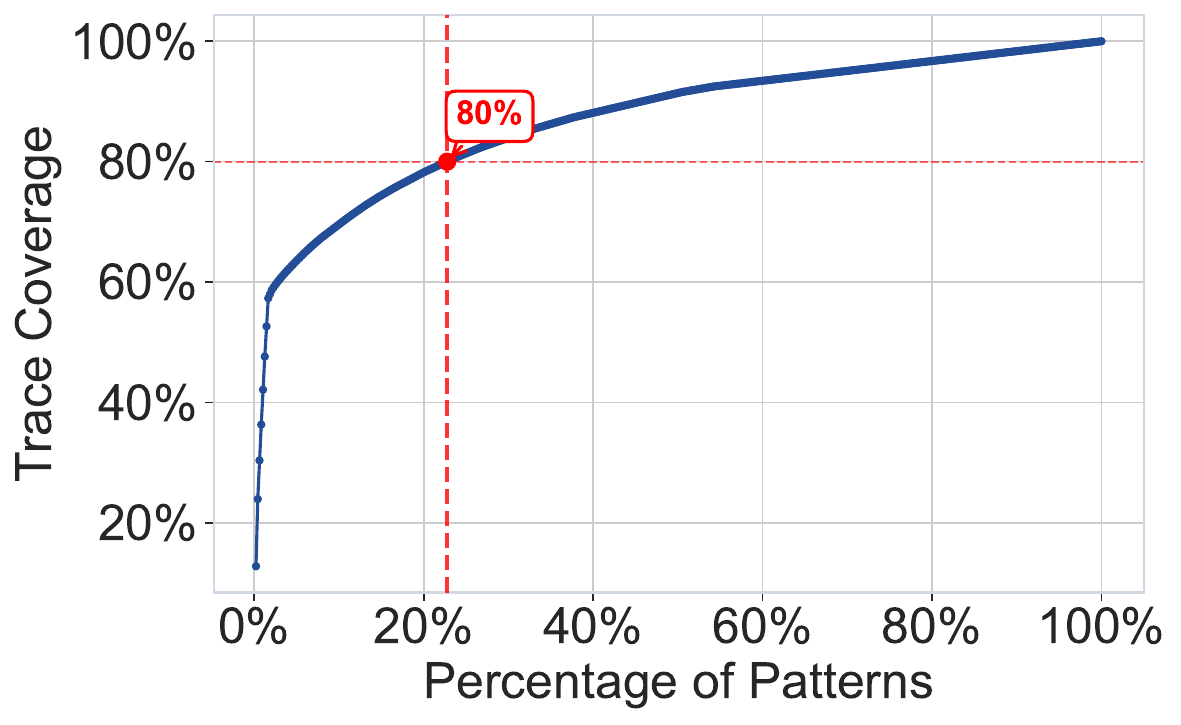}
    \caption{Cumulative coverage of identified invocation patterns in the Comment service: a small number of dominant patterns capture the majority of traffic.}
    \label{fig:pattern_coverage_comment}
\end{figure}

To ensure comprehensive coverage of temporal dynamics, we collected one-week traces from randomly sampled time windows in production clusters, totaling over 510,000 distributed traces. Notably, services in the comment domain generate highly complex execution paths: for a representative critical service, each trace contains an average of over 1,000 gRPC spans, with call depths exceeding 10 layers and breadth spanning more than 100 unique services. This complexity poses significant challenges for accurate pattern extraction and clustering. We focus on the Comment service for detailed analysis, using 245,000 traces collected over 7 days with 7 sampling periods per day. Our approach successfully identifies recurring structural patterns in the invocation graphs. As shown in Fig. \ref{fig:pattern_coverage_comment}, only the 8 most frequent patterns account for over 50\% of all observed traces, and the top 110 patterns cover 80\% of the total trace volume. This high degree of pattern concentration validates the feasibility of leveraging representative templates for resource prediction and optimization, rather than treating each trace as unique.

In addition to accuracy, we assess computational efficiency under heavy load. For traces containing over 1,000 spans, the average graph construction time is 0.3 seconds per 1,000 traces, and pattern matching completes within 1.2 seconds per 1,000 traces on average. These latencies are sufficiently low to support near real-time deployment in production environments, where trace ingestion occurs continuously at scale.

\begin{table}[htbp]
\small
\centering
\caption{Resource demand prediction accuracy comparison across models (lower MAE and higher  $ R^2 $  indicate better performance).}
\label{tab:model_acc_comparison}
\begin{tabular}{lccc}
\toprule
\textbf{Model} & \textbf{MAE} & \textbf{CPU}  $ R^2 $  & \textbf{Memory}  $ R^2 $  \\
\midrule
Linear Regression & 0.0649 & 0.4889 & 0.3681 \\
LSTM & 0.0634 & 0.4956 & 0.3228 \\
DNN & 0.0512 & 0.7531 & 0.5693 \\
MLP & 0.0462 & 0.7813 & 0.6135 \\ 
Random Forest & 0.0251 & 0.9074 & 0.9128 \\
\textbf{Morphis} & \textbf{0.0217} & \textbf{0.9397} & \textbf{0.9367} \\
\bottomrule
\end{tabular}
\end{table}

\subsection{Resource Demand Prediction Accuracy}
We compare the resource demand prediction performance of Morphis against a diverse set of baseline models: Linear Regression, LSTM, DNN, MLP, and Random Forest. These represent canonical approaches across four categories: (i) classical regression (Linear Regression), (ii) deep feedforward networks (DNN, MLP), (iii) sequential modeling (LSTM), and (iv) ensemble tree methods (Random Forest). This selection enables a comprehensive comparison across model families with varying capacities for capturing non-linear and temporal dependencies.

Morphis's prediction model leverages a rich set of multi-dimensional features critical to microservice performance, including runtime metrics (QPS, 95th/99th percentile latency), system state (CPU, memory, and network utilization at the pod level), and concurrency dynamics (request concurrency and QPS over the past 5-second sliding window). To model the complex, non-linear relationship between SLO targets and resource demands, we design a deep neural network architecture built upon multiple residual blocks, which facilitates training stability and enhances representational power. For fairness, we perform extensive hyperparameter tuning for each baseline model by optimizing network depth, hidden dimensions, learning rates, and ensemble sizes to ensure they operate at peak performance. As shown in Table~\ref{tab:model_acc_comparison}, Morphis achieves the lowest Mean Absolute Error (MAE) of 0.0217, outperforming the second-best method (Random Forest, MAE: 0.0251) by 13.6\%. It also attains the highest coefficient of determination ($R^2$) for both CPU (0.9397) and memory (0.9367) demand prediction, indicating superior fit to observed resource consumption. These results demonstrate that our architecture effectively captures the intricate interactions among service behavior, system load, and resource requirements, enabling highly accurate resource prediction.

\subsection{Accuracy of Workload Prediction}
\begin{figure}[htbp]
    \centering
    \subfloat[Forecast comparison over time\label{fig:forecast1}]{%
        \includegraphics[width=0.47\linewidth]{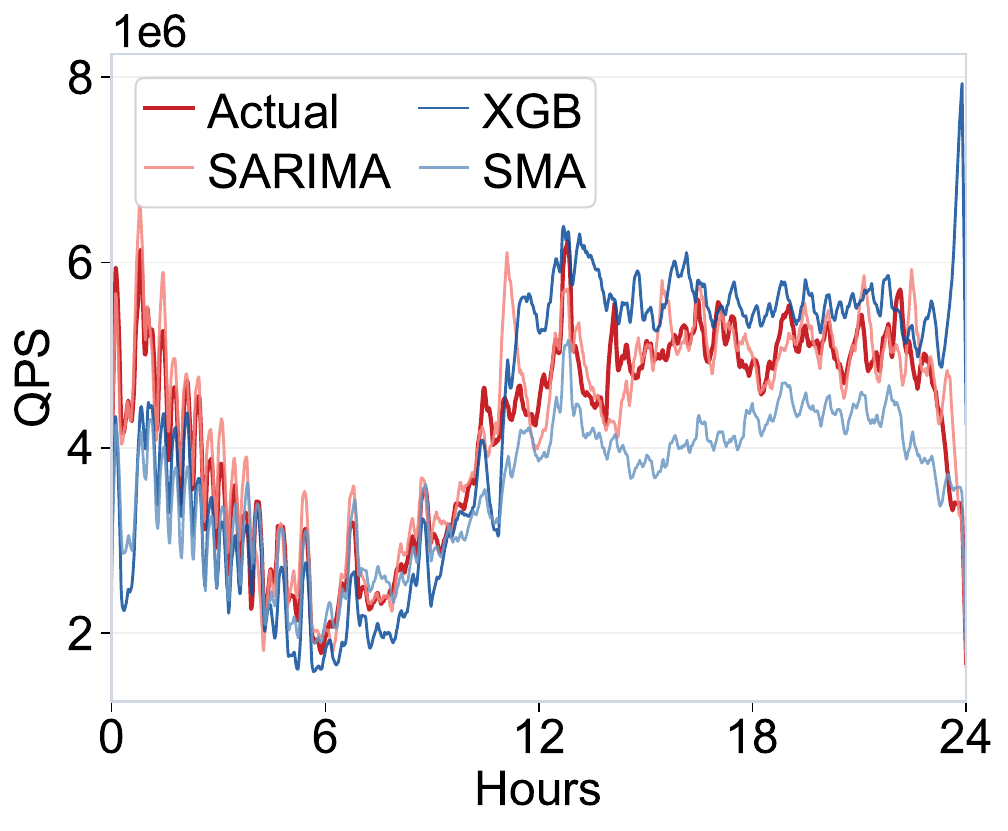}%
    }%
    \subfloat[Model performance comparison\label{fig:forecast2}]{%
        \includegraphics[width=0.47\linewidth]{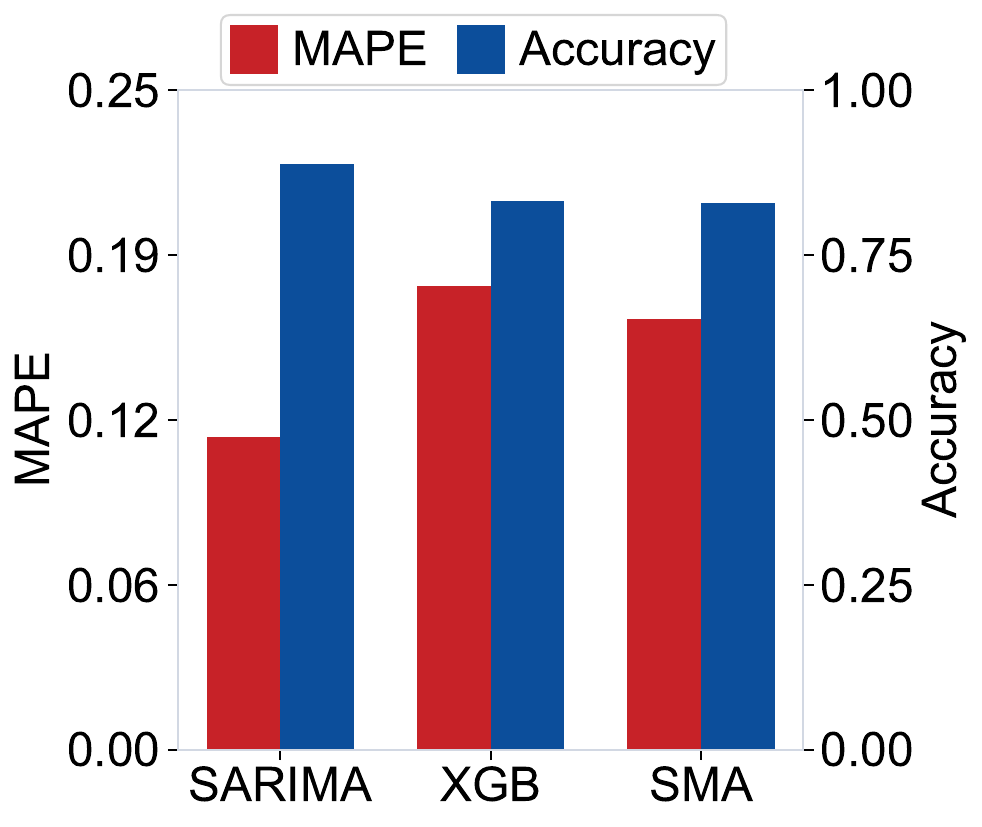}%
    }
    \caption{Workload prediction performance across different methods.}
    \label{fig:forecast}
\end{figure}

Fig. \ref{fig:forecast} evaluates our SARIMA-based approach against XGBoost and seasonal moving average (SMA) baselines. As shown in Fig. \ref{fig:forecast1}, our model accurately tracks the actual workload over a 24-hour period, capturing both diurnal patterns and transient spikes (e.g., during hours 12-15). In contrast, XGBoost tends to overpredict during peaks, while SMA exhibits significant lag due to its reactive nature.
Fig. \ref{fig:forecast2} quantifies the performance gap: our approach achieves a Mean Absolute Percentage Error (MAPE) of 11.5\% and an accuracy of 89\%, representing a 36\% reduction in MAPE compared to XGBoost (18.1\%) and a substantial improvement over SMA (16.7\%). This gain stems from our explicit modeling of temporal structure through weekday and hour-of-day features, which proves more effective than black-box machine learning methods that tend to overfit given the limited training data typical in online production environments.

\subsection{End-to-End Validation}
To evaluate Morphis under realistic deployment conditions, we conducted end-to-end experiments on a production-grade Kubernetes cluster using the \texttt{TrainTicket} microservice benchmark as the target system. We first analyzed multi-day production traces to identify common temporal load patterns, such as gradual ramps, sustained peaks, and transient spikes. Based on this analysis, we designed three representative load scenarios (see Fig. \ref{fig:exp_load_shape}): S1 (Gradual Rise), S2 (High Sustained Load), and S3 (Spike-and-Plateau). Each experiment runs for one hour, with traffic patterns generated using Locust to closely emulate the shape and dynamics of real-world production workloads.\footnote{Locust is an open source load testing tool. See \url{https://locust.io/} for details.} While the full-scale production traces are too large to replay directly in our testbed, we manually calibrated the Locust scripts to reproduce the key characteristics (e.g., request rate, burstiness, duration) observed in the original traces. Five independent trials are conducted per configuration to reduce variance and improve result reliability.

\begin{figure}[t]
\centering
\subfloat[Scenario 1: Gradual Rise\label{fig:cpu_quota_by_service_s1}]{\includegraphics[width=\linewidth]{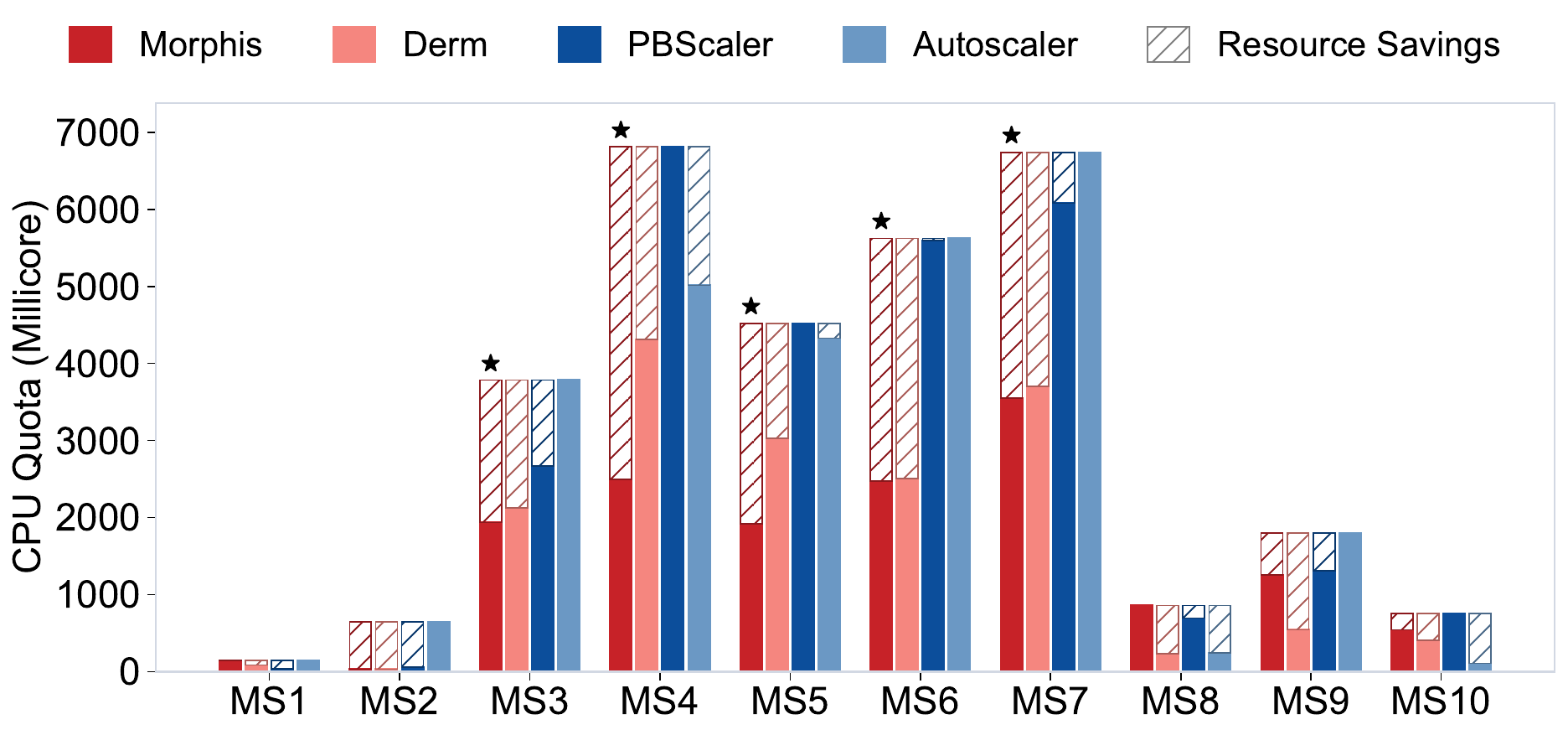}}\\
\subfloat[Scenario 2: High Sustained\label{fig:cpu_quota_by_service_s2}]{\includegraphics[width=\linewidth]{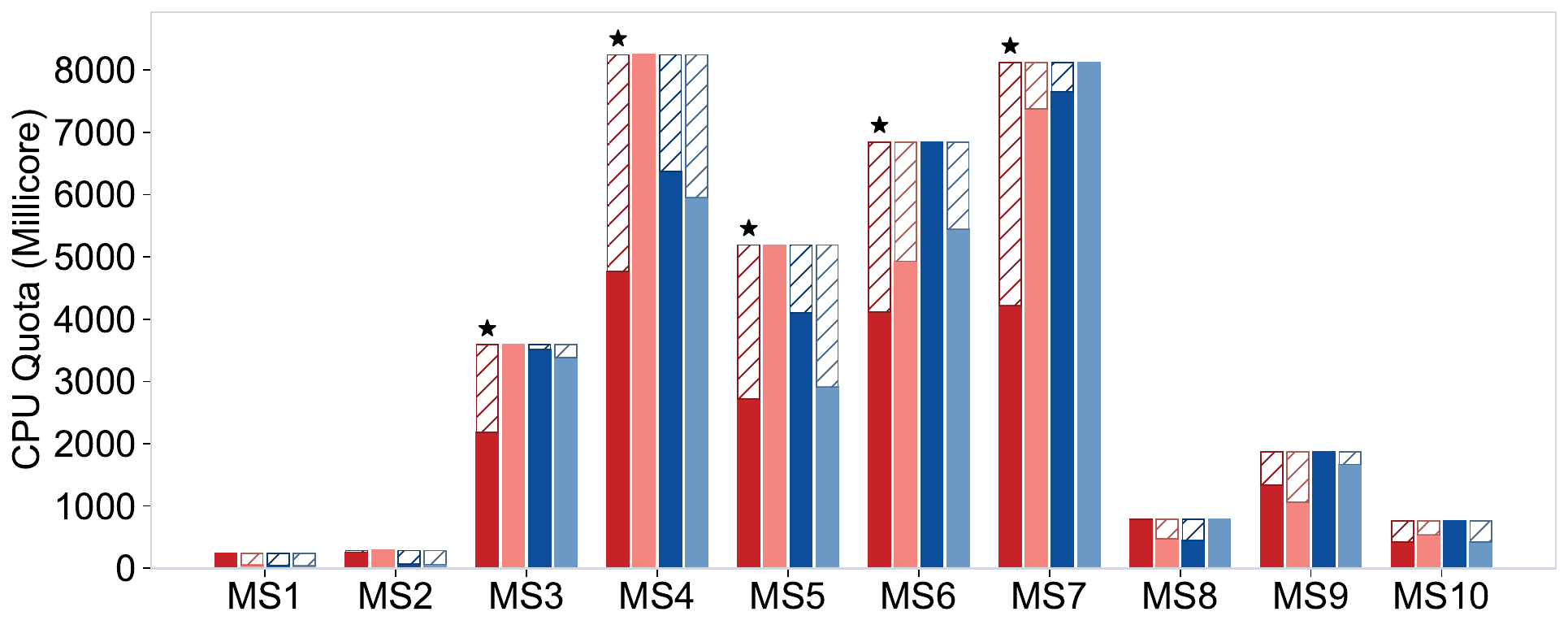}}\\
\subfloat[Scenario 3: Spike and Plateau\label{fig:cpu_quota_by_service_s3}]{\includegraphics[width=\linewidth]{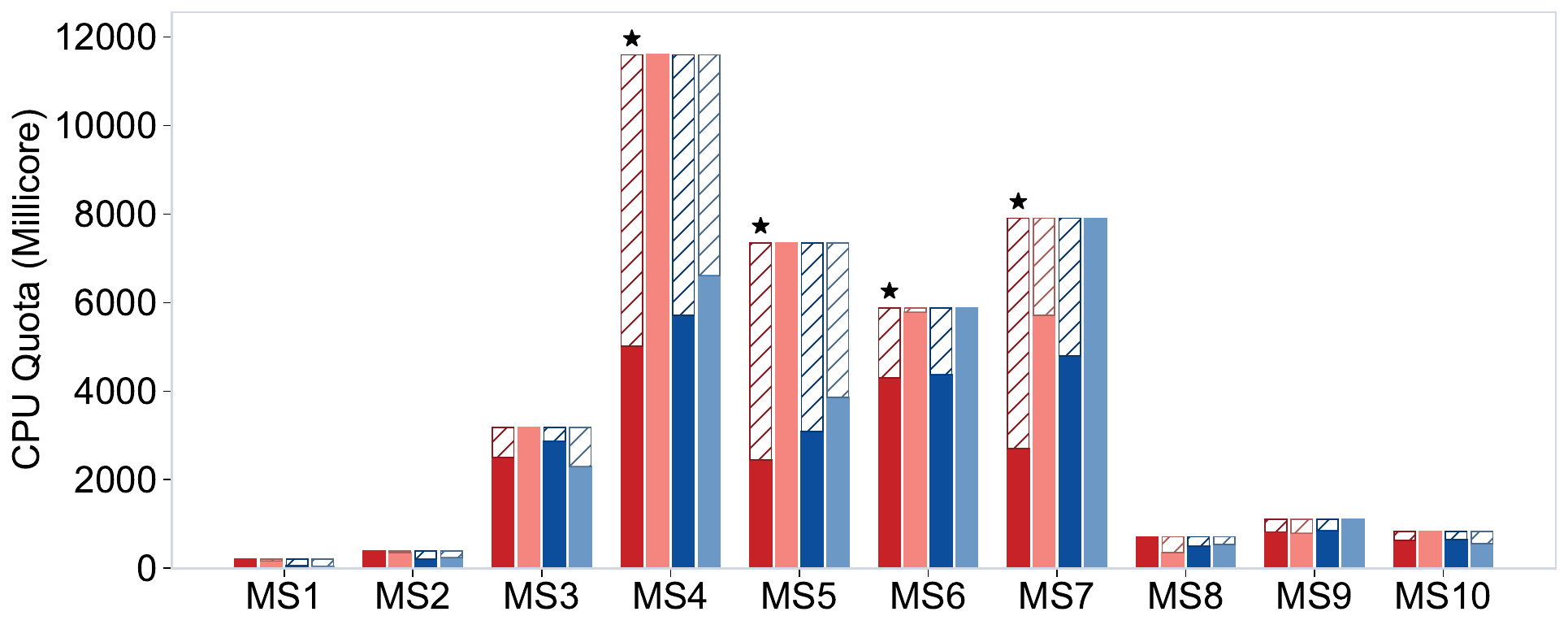}}
\caption{Per-microservice CPU allocation (in CPU cores) under different load patterns. MS1-MS10 denote ten major microservices in \texttt{TrainTicket}; bars represent average allocation.}
\label{fig:cpu_consumption_per_service}
\end{figure}

\begin{figure}[htbp]
    \centering
    \includegraphics[width=1\linewidth]{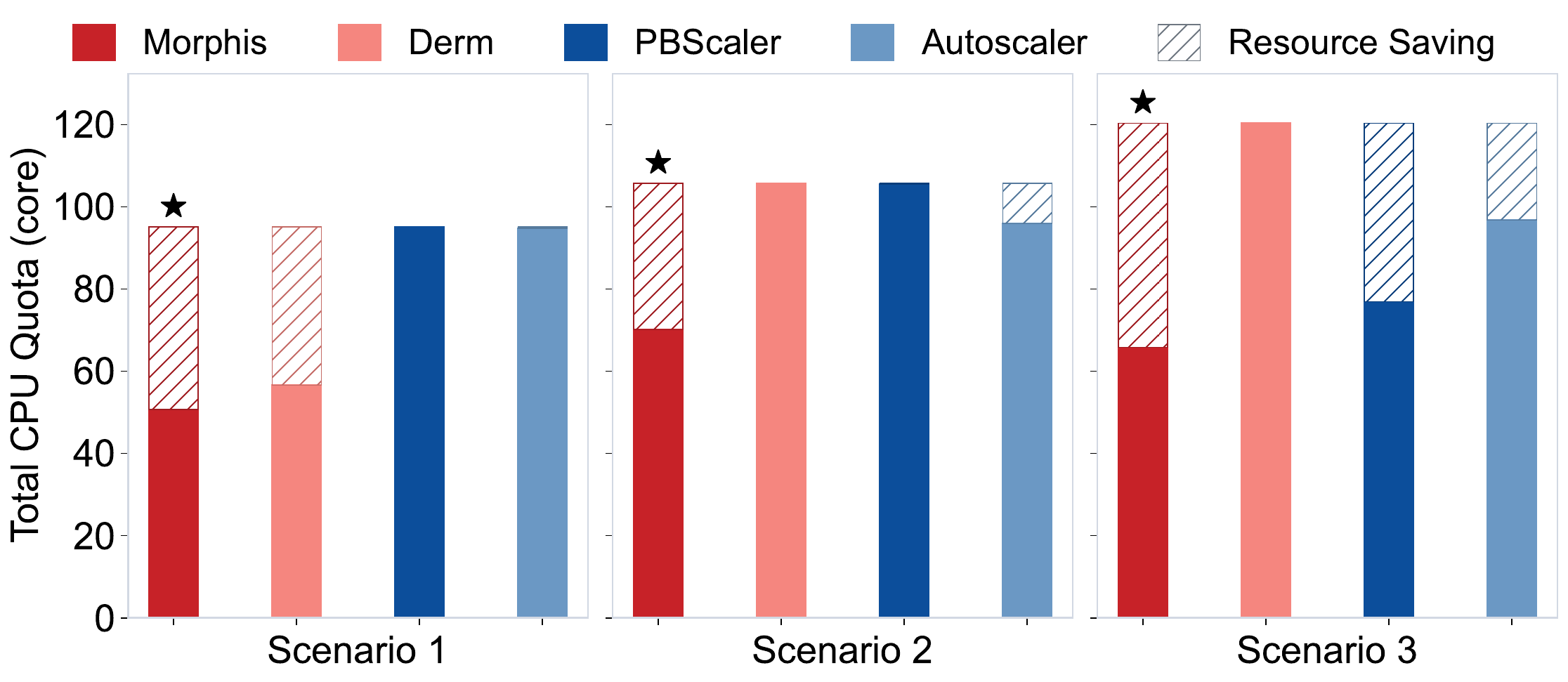}
    \caption{Total CPU consumption across all major microservices. Morphis reduces aggregate resource usage by up to 38\% compared to HPA, while maintaining SLO compliance.}
    \label{fig:total_cpu_consumption}
\end{figure}

\subsubsection{Resource Efficiency}
Morphis achieves superior resource efficiency across all scenarios. As shown in Fig. \ref{fig:total_cpu_consumption}, it reduces total CPU consumption by up to 38\% compared to the next best method. Under the most challenging High Sustained Load pattern (S2), Morphis delivers an additional 37.5\% resource savings over existing approaches. At the service level, Morphis achieves optimal utilization in 7 out of 10 core microservices, with peak savings exceeding 50\% in several bottleneck services (see Fig. \ref{fig:cpu_consumption_per_service}). This stems from its ability to perform fine-grained, inter-service coordination through global optimization rather than isolated per-service scaling.

\subsubsection{Latency Performance and SLO Compliance}
\begin{figure}[htbp]
    \centering
    \includegraphics[width=1\linewidth]{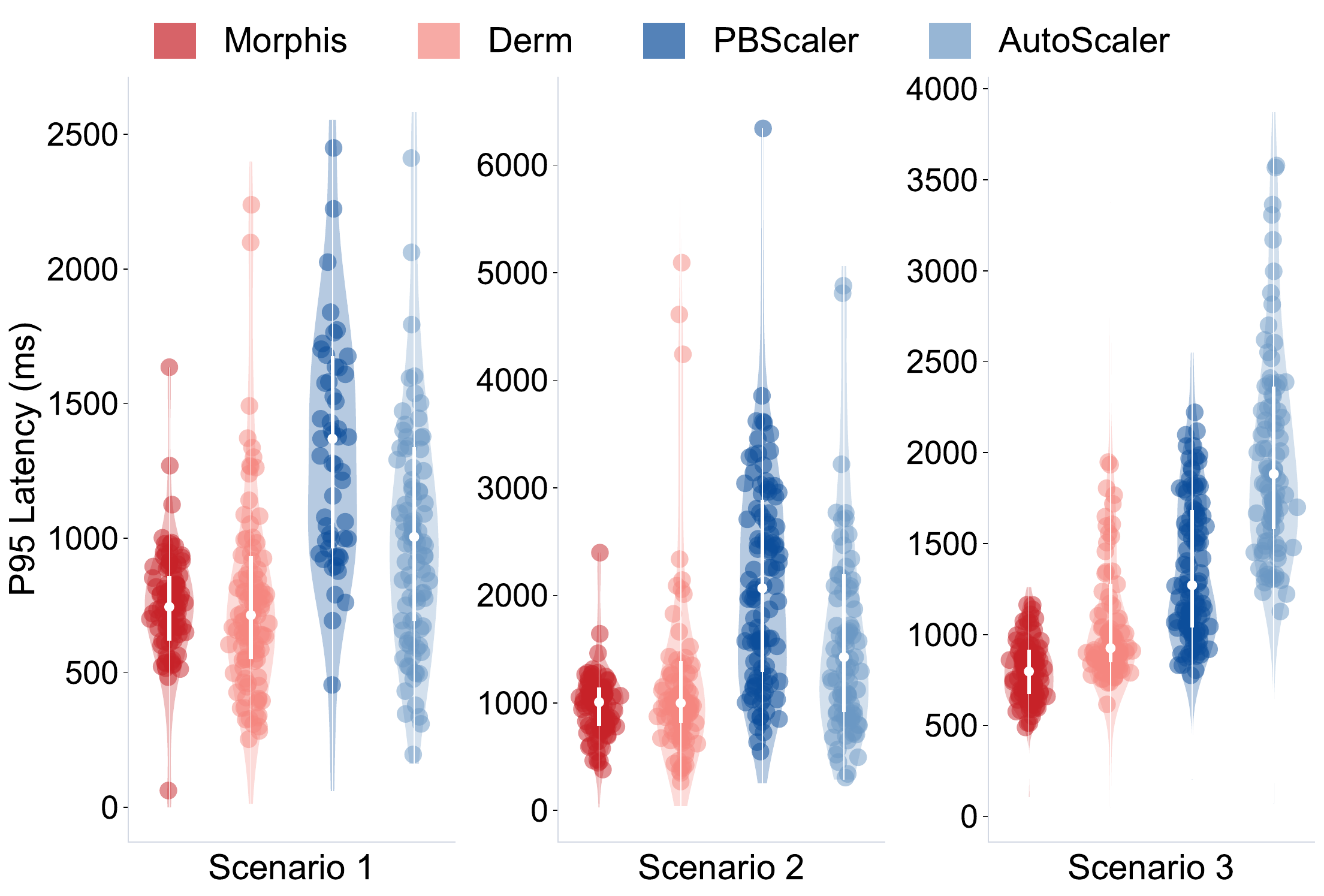}
    \caption{P95 latency distribution across methods. Morphis shows tighter, lower-latency concentration.}
    \label{fig:latency_violin}
\end{figure}

The High Sustained Load scenario poses the greatest challenge due to rapid demand fluctuations and high concurrency, requiring fast and accurate scaling responses to avoid latency degradation. Latency results, summarized in Fig. \ref{fig:latency_violin}, show that Morphis maintains the lowest and most stable P95 latency across all scenarios.

Latency distributions follow a characteristic ``short and wide'' shape in the violin plots, indicating tight concentration around low values with minimal tail latency. This reflects the effectiveness of proactive scaling guided by invocation pattern-aware predictions. We define SLO targets based on acceptable user experience: {S1: 1000\,ms}, {S2: 1200\,ms}, {S3: 1000\,ms}. Contrary to typical reporting of violation rates, we report the \textit{SLO compliance rate}, i.e., the percentage of requests meeting latency targets. Morphis achieves high SLO compliance: 98.80\% (S1), 96.67\% (S2), and 95.57\% (S3) under Gradual Rise, High Sustained, and Spike workloads, respectively. These results demonstrate that Morphis not only minimizes resource usage but also ensures high responsiveness under dynamic loads.

\subsubsection{Global Optimization Mechanism}
Unlike reactive or per-service baselines, Morphis performs end-to-end global optimization across the entire service mesh, dynamically coordinating resource allocation in response to evolving invocation patterns and performance constraints. A key innovation is its ability to \textit{adaptively adjust per-microservice latency targets} based on real-time bottleneck analysis: it tightens latency objectives in non-critical (non-bottleneck) services to reduce unnecessary over-provisioning, while conservatively relaxing them in constrained paths to prevent resource starvation and cascading delays. This dynamic rebalancing ensures end-to-end SLO compliance with minimal total CPU allocation, enabling coordinated, system-wide efficiency.

\begin{figure}[t]
    \centering
    \includegraphics[width=0.85\linewidth]{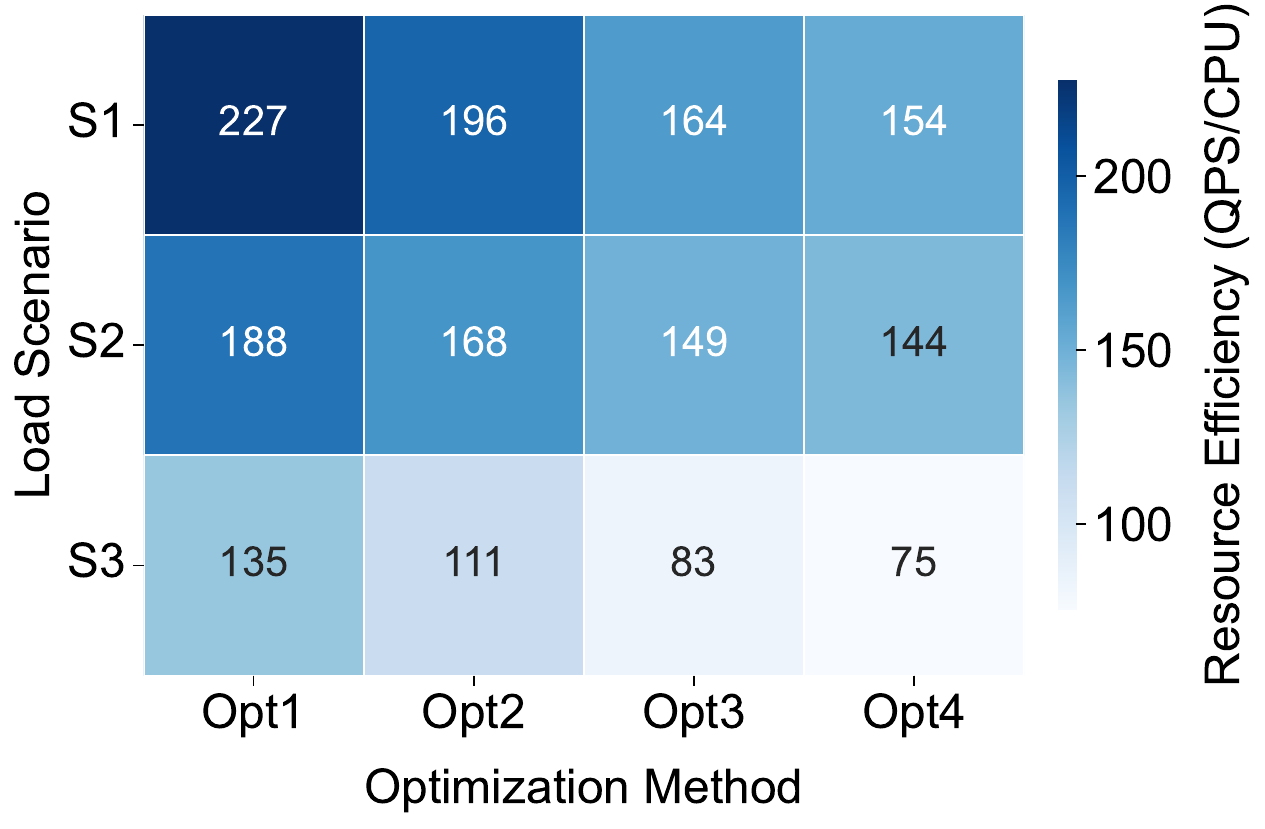}
    \caption{Resource efficiency across three load scenarios and four optimization strategies. We report the QPS-per-CPU ratio as a measure of cost-effectiveness.}
    \label{fig:opt_heatmap}
\end{figure}

To isolate the impact of our global optimization design, we conduct an ablation study comparing four variants:
\begin{itemize}
    \item \textit{Opt1 (Morphis)}: Global optimization with dynamic dependency tracking,
    \item \textit{Opt2}: Dynamic call graph but \textit{independent} per-service scaling,
    \item \textit{Opt3}: Static call graph with global optimization,
    \item \textit{Opt4}: Independent scaling under static dependencies (fully decentralized).
\end{itemize}
As shown in Fig. \ref{fig:opt_heatmap}, which reports the QPS-per-CPU ratio as a measure of cost-effectiveness, \textit{Opt1} consistently achieves the highest performance across all three load scenarios. Compared to the static, decentralized baseline (\textit{Opt4}), Morphis improves efficiency by 47.8\% (S1), 30.3\% (S2), and 78.9\% (S3). Even against the dynamic but non-coordinated \textit{Opt2}, gains reach 15.8\% (S1), 11.6\% (S2), and 21.7\% (S3), confirming the value of global coordination. Notably, the largest improvement occurs in S3 (Spike and Plateau), where workload shifts are abrupt yet predictable. This highlights Morphis's strength in leveraging temporal regularity: by proactively adjusting service-level objectives in anticipation of emerging bottlenecks, it avoids reactive over-scaling and maintains high efficiency throughout the transition phase. In contrast, methods without a global view (e.g., Opt2, Opt4) fail to coordinate scaling actions, leading to resource contention and suboptimal allocation.

In summary, Morphis achieves both higher resource efficiency and stronger reliability than prior approaches, demonstrating the practical benefits of pattern-aware, globally optimized scheduling in dynamic microservice environments.

\section{Related Work}
\label{sec:related}

%%% TODO for ty：看看是否有关于 microservice trace compression 的工作，好像是发在ATC23上
% 另外可以见缝插针引用如下我们自己的工作：
% 1. Distributed redundant placement for microservice-based applications at the edge（调度）
% 2. Learning to schedule multi-server jobs with fluctuated processing speeds（调度）
% 3. Data-locality-aware task assignment and scheduling for distributed job executions（调度）

Auto-scaling in microservice systems has evolved along two interdependent dimensions: the sophistication of scaling strategies and the depth of dependency awareness. Early approaches treat services as isolated units, triggering scaling based on local metrics such as CPU utilization or request queue length~\cite{aws_autoscaling, alibaba_autoscaling, google_autoscaler, ibm_autoscaling}. While simple and widely deployed (e.g., Kubernetes HPA~\cite{kubernetes_hpa}), these \textit{reactive} methods suffer from inherent latency, often causing SLO violations during sudden load surges or resource waste during declines. To overcome this, \textit{proactive} methods leverage workload forecasting, using attention networks~\cite{pan2023magicscaler}, meta-reinforcement learning~\cite{xue2022meta}, or burst detection~\cite{abdullah2020burst}, to scale ahead of demand. However, most predictive models operate per-service and ignore how scaling one component propagates effects across the service mesh. Even hybrid reactive-predictive frameworks lack an explicit model of end-to-end request flow, limiting their ability to coordinate resources for global SLO compliance.

Recognizing these limitations, recent work turns to call graph analysis derived from distributed traces to capture inter-service dependencies. Empirical studies reveal that real-world microservice topologies are not arbitrary but exhibit structural regularities: hierarchical motifs~\cite{luo2022depth}, recurring execution paths modeled as DAGs~\cite{wen2022characterizing}, and identifiable critical paths for bottleneck diagnosis~\cite{zhang2022crisp}. These findings confirm that invocation patterns concentrate around stable structures, offering a foundation for smarter orchestration. Yet, much of this research remains descriptive or diagnostic. It characterizes “what happened” but does not translate structural insights into prescriptive control actions.

Building on dependency awareness, a growing line of work integrates call graphs into resource management. Systems like ERMS~\cite{luo2022erms} optimize per-service allocations under global latency constraints; Firm~\cite{qiu2020firm} uses trace mining to target SLO-critical services; Nodens~\cite{shi2023nodens} and FastPert~\cite{xu2025fastpert} enable fine-grained scheduling via load or task decomposition; and dRPC~\cite{bai2024drpc} applies deep reinforcement learning for distributed allocation. Despite these advances, existing solutions typically assume static or slowly evolving dependencies and perform localized optimizations, either scaling bottlenecks in isolation or applying uniform policies across services. They fail to adapt to runtime topology shifts (e.g., from feature flags or conditional logic) and miss opportunities for cross-service trade-offs that could reduce total resource cost while preserving end-to-end SLOs.

In contrast, Morphis bridges these gaps by unifying dynamic pattern discovery, semantic dependency abstraction, and global optimization. It recognizes that despite apparent trace-level chaos, invocation structures concentrate into a small set of recurring \textit{structural fingerprints}. Morphis continuously extracts these fingerprints via backbone-subgraph decomposition and uses them as the basis for a closed-loop optimizer that jointly predicts pattern prevalence and allocates resources across all services under end-to-end SLO constraints. This enables proactive, coordinated provisioning that adapts to evolving dependencies.
\section{Conclusion}\label{sec:conclusion}
In this paper, we present Morphis, a context-aware resource management framework that addresses the critical challenge of dynamic dependencies in microservice architectures. Unlike traditional approaches that rely on static call graphs, 
Morphis captures the evolving nature of service interactions through a novel combination of structured tracing, fingerprint-based pattern matching, and lightweight machine learning. By identifying recurring invocation patterns over time, our system enables proactive resource scheduling. At the core of Morphis is a global optimization framework that dynamically adjusts per-service latency targets based on real-time bottleneck analysis, ensuring end-to-end SLO compliance with minimal resource allocation. This coordinated, dependency-aware approach avoids both over-provisioning in non-critical paths and under-provisioning in emerging bottlenecks. We evaluate Morphis on a Kubernetes cluster using the \texttt{TrainTicket} benchmark under diverse, trace-inspired load scenarios. Experimental results show that Morphis reduces CPU consumption by up to 38\% compared to state-of-the-art baselines while consistently meeting SLOs.

\bibliographystyle{IEEEtran}
\bibliography{references} 

\begin{thebibliography}{10}
\providecommand{\url}[1]{#1}
\csname url@samestyle\endcsname
\providecommand{\newblock}{\relax}
\providecommand{\bibinfo}[2]{#2}
\providecommand{\BIBentrySTDinterwordspacing}{\spaceskip=0pt\relax}
\providecommand{\BIBentryALTinterwordstretchfactor}{4}
\providecommand{\BIBentryALTinterwordspacing}{\spaceskip=\fontdimen2\font plus
\BIBentryALTinterwordstretchfactor\fontdimen3\font minus \fontdimen4\font\relax}
\providecommand{\BIBforeignlanguage}[2]{{%
\expandafter\ifx\csname l@#1\endcsname\relax
\typeout{** WARNING: IEEEtran.bst: No hyphenation pattern has been}%
\typeout{** loaded for the language `#1'. Using the pattern for}%
\typeout{** the default language instead.}%
\else
\language=\csname l@#1\endcsname
\fi
#2}}
\providecommand{\BIBdecl}{\relax}
\BIBdecl

\bibitem{zhang2025bridge}
T.~Zhang, Y.~Liang, G.~Li, T.~Tan, C.~Xu, and Y.~Li, ``Bridge the islands: Pointer analysis for microservice systems,'' \emph{Proceedings of the ACM on Software Engineering}, vol.~2, no. ISSTA, pp. 504--526, 2025.

\bibitem{rzadca2020autopilot}
K.~Rzadca, P.~Findeisen, J.~Swiderski, P.~Zych, P.~Broniek, J.~Kusmierek, P.~Nowak, B.~Strack, P.~Witusowski, S.~Hand \emph{et~al.}, ``Autopilot: workload autoscaling at google,'' in \emph{Proceedings of the Fifteenth European Conference on Computer Systems}, 2020, pp. 1--16.

\bibitem{zhang2022crisp}
Z.~Zhang, M.~K. Ramanathan, P.~Raj, A.~Parwal, T.~Sherwood, and M.~Chabbi, ``Crisp: Critical path analysis of large-scale microservice architectures,'' in \emph{2022 USENIX Annual Technical Conference (USENIX ATC 22)}, 2022, pp. 655--672.

\bibitem{huye2023lifting}
D.~Huye, Y.~Shkuro, and R.~R. Sambasivan, ``Lifting the veil on meta’s microservice architecture: Analyses of topology and request workflows,'' in \emph{2023 USENIX Annual Technical Conference (USENIX ATC 23)}, 2023, pp. 419--432.

\bibitem{lu2017imbalance}
C.~Lu, K.~Ye, G.~Xu, C.-Z. Xu, and T.~Bai, ``Imbalance in the cloud: An analysis on alibaba cluster trace,'' in \emph{2017 IEEE International Conference on Big Data (Big Data)}.\hskip 1em plus 0.5em minus 0.4em\relax IEEE, 2017, pp. 2884--2892.

\bibitem{hou2019unleashing}
X.~Hou, J.~Liu, C.~Li, and M.~Guo, ``Unleashing the scalability potential of power-constrained data center in the microservice era,'' in \emph{Proceedings of the 48th International Conference on Parallel Processing}, 2019, pp. 1--10.

\bibitem{he2022online}
X.~He, Z.~Tu, M.~Wagner, X.~Xu, and Z.~Wang, ``Online deployment algorithms for microservice systems with complex dependencies,'' \emph{IEEE Transactions on Cloud Computing}, vol.~11, no.~2, pp. 1746--1763, 2022.

\bibitem{luo2022erms}
S.~Luo, H.~Xu, K.~Ye, G.~Xu, L.~Zhang, J.~He, G.~Yang, and C.~Xu, ``Erms: Efficient resource management for shared microservices with sla guarantees,'' in \emph{Proceedings of the 28th ACM International Conference on Architectural Support for Programming Languages and Operating Systems, Volume 1}, 2022, pp. 62--77.

\bibitem{li2023topology}
X.~Li, J.~Zhou, X.~Wei, D.~Li, Z.~Qian, J.~Wu, X.~Qin, and S.~Lu, ``Topology-aware scheduling framework for microservice applications in cloud,'' \emph{IEEE Transactions on Parallel and Distributed Systems}, vol.~34, no.~5, pp. 1635--1649, 2023.

\bibitem{ekane2025disc}
B.~Ekane, D.~Mvondo, R.~Lachaize, Y.-D. Bromberg, A.~Tchana, and D.~Hagimont, ``Disc: Backpressure mitigation in multi-tier applications with distributed shared connection,'' in \emph{22nd USENIX Symposium on Networked Systems Design and Implementation (NSDI 25)}, 2025, pp. 55--70.

\bibitem{gan2019open}
Y.~Gan, Y.~Zhang, D.~Cheng, A.~Shetty, P.~Rathi, N.~Katarki, A.~Bruno, J.~Hu, B.~Ritchken, B.~Jackson \emph{et~al.}, ``An open-source benchmark suite for microservices and their hardware-software implications for cloud \& edge systems,'' in \emph{Proceedings of the twenty-fourth international conference on architectural support for programming languages and operating systems}, 2019, pp. 3--18.

\bibitem{qiu2020firm}
H.~Qiu, S.~S. Banerjee, S.~Jha, Z.~T. Kalbarczyk, and R.~K. Iyer, ``Firm: An intelligent fine-grained resource management framework for slo-oriented microservices,'' in \emph{14th USENIX symposium on operating systems design and implementation (OSDI 20)}, 2020, pp. 805--825.

\bibitem{Istio}
\BIBentryALTinterwordspacing
Istio. (2025) Istio service mesh. [Online]. Available: \url{https://istio.io/}
\BIBentrySTDinterwordspacing

\bibitem{zhang2017residual}
K.~Zhang, M.~Sun, T.~X. Han, X.~Yuan, L.~Guo, and T.~Liu, ``Residual networks of residual networks: Multilevel residual networks,'' \emph{IEEE Transactions on Circuits and Systems for Video Technology}, vol.~28, no.~6, pp. 1303--1314, 2017.

\bibitem{Jaeger}
{Jaeger Authors}, ``{Jaeger},'' \url{https://www.jaegertracing.io/}, 2025, accessed: 2025-09-03.

\bibitem{Prometheus}
{Prometheus Authors}, ``{Prometheus},'' \url{https://prometheus.io/}, 2025, accessed: 2025-09-03.

\bibitem{zhou2018benchmarking}
X.~Zhou, X.~Peng, T.~Xie, J.~Sun, C.~Xu, C.~Ji, and W.~Zhao, ``Benchmarking microservice systems for software engineering research,'' in \emph{Proceedings of the 40th International Conference on Software Engineering: Companion Proceeedings}, 2018, pp. 323--324.

\bibitem{chen2024derm}
L.~Chen, S.~Luo, C.~Lin, Z.~Mo, H.~Xu, K.~Ye, and C.~Xu, ``Derm: Sla-aware resource management for highly dynamic microservices,'' in \emph{2024 ACM/IEEE 51st Annual International Symposium on Computer Architecture (ISCA)}.\hskip 1em plus 0.5em minus 0.4em\relax IEEE, 2024, pp. 424--436.

\bibitem{xie2024pbscaler}
S.~Xie, J.~Wang, B.~Li, Z.~Zhang, D.~Li, and P.~C. Hung, ``Pbscaler: A bottleneck-aware autoscaling framework for microservice-based applications,'' \emph{IEEE Transactions on Services Computing}, vol.~17, no.~2, pp. 604--616, 2024.

\bibitem{kubernetes_hpa}
\BIBentryALTinterwordspacing
kubernetes. (2025) Kubernetes horizontal pod autoscaling. [Online]. Available: \url{https://kubernetes.io/docs/tasks/run-application/horizontal-pod-autoscale/}
\BIBentrySTDinterwordspacing

\bibitem{aws_autoscaling}
{Amazon Web Services}, ``{Amazon EC2 Auto Scaling User Guide},'' \url{https://docs.aws.amazon.com/autoscaling/}, 2025, accessed: 2025-09-03.

\bibitem{alibaba_autoscaling}
{Alibaba Cloud}, ``{Auto Scaling},'' \url{https://www.alibabacloud.com/en/product/auto-scaling}, 2025, accessed: 2025-09-03.

\bibitem{google_autoscaler}
{Google Cloud}, ``{Compute Engine Autoscaler},'' \url{https://cloud.google.com/compute/docs/autoscaler}, 2025, accessed: 2025-09-03.

\bibitem{ibm_autoscaling}
{IBM}, ``{Horizontal pod auto scaling},'' \url{https://www.ibm.com/docs/en/order-management-sw/10.0.0?topic=SS6PEW_10.0.0/installation/t_OMICP_horiz_pod_autoscale.htm}, 2025, accessed: 2025-09-03.

\bibitem{pan2023magicscaler}
Z.~Pan, Y.~Wang, Y.~Zhang, S.~B. Yang, Y.~Cheng, P.~Chen, C.~Guo, Q.~Wen, X.~Tian, Y.~Dou \emph{et~al.}, ``Magicscaler: Uncertainty-aware, predictive autoscaling,'' \emph{Proceedings of the VLDB Endowment}, vol.~16, no.~12, pp. 3808--3821, 2023.

\bibitem{xue2022meta}
S.~Xue, C.~Qu, X.~Shi, C.~Liao, S.~Zhu, X.~Tan, L.~Ma, S.~Wang, S.~Wang, Y.~Hu \emph{et~al.}, ``A meta reinforcement learning approach for predictive autoscaling in the cloud,'' in \emph{Proceedings of the 28th ACM SIGKDD Conference on Knowledge Discovery and Data Mining}, 2022, pp. 4290--4299.

\bibitem{abdullah2020burst}
M.~Abdullah, W.~Iqbal, J.~L. Berral, J.~Polo, and D.~Carrera, ``Burst-aware predictive autoscaling for containerized microservices,'' \emph{IEEE Transactions on Services Computing}, vol.~15, no.~3, pp. 1448--1460, 2020.

\bibitem{luo2022depth}
S.~Luo, H.~Xu, C.~Lu, K.~Ye, G.~Xu, L.~Zhang, J.~He, and C.~Xu, ``An in-depth study of microservice call graph and runtime performance,'' \emph{IEEE Transactions on Parallel and Distributed Systems}, vol.~33, no.~12, pp. 3901--3914, 2022.

\bibitem{wen2022characterizing}
Y.~Wen, G.~Cheng, S.~Deng, and J.~Yin, ``Characterizing and synthesizing the workflow structure of microservices in bytedance cloud,'' \emph{Journal of Software: Evolution and Process}, vol.~34, no.~8, p. e2467, 2022.

\bibitem{shi2023nodens}
J.~Shi, H.~Zhang, Z.~Tong, Q.~Chen, K.~Fu, and M.~Guo, ``Nodens: Enabling resource efficient and fast qos recovery of dynamic microservice applications in datacenters,'' in \emph{2023 USENIX Annual Technical Conference (USENIX ATC 23)}, 2023, pp. 403--417.

\bibitem{xu2025fastpert}
H.~Xu, Y.~Liu, S.~Xie, W.~C. Lau \emph{et~al.}, ``Fastpert: Towards fast microservice application latency prediction via structural inductive bias over pert networks,'' in \emph{Proceedings of the AAAI Conference on Artificial Intelligence}, vol.~39, no.~19, 2025, pp. 20\,787--20\,795.

\bibitem{bai2024drpc}
H.~Bai, M.~Xu, K.~Ye, R.~Buyya, and C.~Xu, ``Drpc: Distributed reinforcement learning approach for scalable resource provisioning in container-based clusters,'' \emph{IEEE Transactions on Services Computing}, 2024.

\end{thebibliography}

\end{document}